\documentclass[]{aa} 
\usepackage{txfonts}
\usepackage{natbib}
\usepackage{graphicx, xspace}
\usepackage{placeins}
\usepackage{array}
\usepackage[switch]{lineno}
\usepackage{hyperref}
\usepackage{color}
\usepackage{gensymb}

\newcommand{\Lagr}{\mathcal{L}}
\newcommand{\Magr}{\mathcal{M}}

\makeatletter

\newcommand{\Rmnum}[1]{\expandafter\@slowromancap\romannumeral #1@}
\makeatother

\def\d{\,d$^{-1}$\xspace}
\begin{document}


\title{Magnetic characterization of the SPB/$\mathbf{\beta}$\,Cep hybrid pulsator HD\,43317\thanks{This work was based on data gathered with HARPS installed on the 3.6-m ESO telescope (ESO Large Programme 182.D-0356) at La Silla, Chile.}$^{,}$\thanks{Mode identification results obtained with the software package FAMIAS developed in the framework of the FP6 European Coordination Action HELAS (http://www.helas-eu.org).}}

\authorrunning{B.~Buysschaert}
\titlerunning{Magnetic characterization of HD\,43317}

\author{B.\,Buysschaert\inst{1,2}, 
C.\,Neiner\inst{1},
M.\,Briquet\inst{1} \and
C.\,Aerts\inst{2,3} 
}
\mail{bram.buysschaert@obspm.fr} 
\date{Received: 20 April 2017 / Accepted: 7 June 2017} 

\institute{
LESIA, Observatoire de Paris, PSL Research University, CNRS, Sorbonne Universit\'es, UPMC Univ. Paris 06, Univ. Paris Diderot, Sorbonne Paris Cit\'e, 5 place Jules Janssen, F-92195 Meudon, France 
\and Instituut voor Sterrenkunde, KU Leuven, Celestijnenlaan 200D, 3001 Leuven, Belgium 
\and Dept. of Astrophysics, IMAPP, Radboud University Nijmegen, 6500 GL, Nijmegen, The Netherlands 
}
\abstract
{Large-scale magnetic fields at the surface of massive stars do not only influence the outer-most layers of the star, but also have consequences for the deep interior, only observationally accessible through asteroseismology.  We performed a detailed characterization of the dipolar magnetic field at the surface of the B3.5V star HD\,43317, a SPB/$\beta$\,Cep hybrid pulsator, by studying the rotationally modulated magnetic field of archival and new Narval spectropolarimetry.  Additionally, we employed a grid-based approach to compare the Zeeman signatures with model profiles.  By studying the rotational modulation of the He lines in both the Narval and HARPS spectroscopy caused by co-rotating surface abundance inhomogeneities, we updated the rotation period to $0.897673\pm0.000004$\,d.  The inclination angle between the rotation axis and the observer's line of sight remains ill-defined, because of the low level of variability in Stokes\,V and deformations in the intensity profiles by stellar pulsation modes.  The obliquity angle between the rotation and magnetic axes is constrained to $\beta \in [67,90]\,^{\circ}$, and the strength of the dipolar magnetic field is of the order of 1\,kG to 1.5\,kG.  This magnetic field at the stellar surface is sufficiently strong to warrant a uniformly rotating radiative envelope, causing less convective core overshooting, which should be visible in future forward seismic modeling.}


\maketitle

\section{Introduction}
\label{sec:Introduction}
\subsection{The effect of magnetism in early-type stars}
Over the last two decades, magnetic fields have been steadily detected for massive stars with spectral type earlier than B3.  This is in large part thanks to the detailed spectropolarimetric surveys that systematically search for such magnetic fields in carefully selected samples that depend on their respective objectives (MiMeS, \citet{2016MNRAS.456....2W}; BinaMIcS, \citet{2015IAUS..307..330A}; the BOB campaign, \citet{2015IAUS..307..342M}; and the BRITE spectropolarimetric survey, \citet{2016arXiv161103285N}).  In other cases, magnetic fields were discovered from indirect observational evidence, such as rotational modulation of X-ray emission, UV resonance lines or H$\alpha$ emission.

Almost all detected magnetic fields at the surface of massive stars have a simple geometry, that is a magnetic dipole inclined to the rotation axis.  These large-scale magnetic fields are of fossil origin.  They are observed to be stable over a time span of decades and their properties do not correlate with known stellar parameters.  As such, they must have been generated during the star's formation phase \citep{2015IAUS..305...61N}.  Such large-scale magnetic fields are detected at the surface of about 10\,\% of the early-type stars \citep[e.g.,][]{2015IAUS..305...53G}, indicating a strong stabilizing mechanism of the earlier generated magnetic fields during the contraction phase towards core-hydrogen burning.  In addition, small-scale dynamo magnetic fields produced in the convective sub-surface layer have been suggested to reach the stellar surface from a theoretical viewpoint \citep[e.g.,][]{2011A+A...534A.140C}.  Yet, no observational evidence has been found so far to corroborate this idea.  Lastly, a dynamo field is expected to be produced by the convective core \citep{1989MNRAS.236..629M}.  However, no direct observational diagnostics at the stellar surface of this field are expected, since the timescale to transport the magnetic flux from the core to the surface is longer than the stellar lifetime.

The large sample of known magnetic early-type stars now permits us to begin to observationally unravel the interactions between the large-scale magnetic field, the star, and its nearby environment.  One of these effects is the influence of the magnetic field on the diffusion and stratification of chemical elements.  The chemically peculiar Ap/Bp stars have their distinct chemical surface abundance to this particular effect \citep[e.g.,][]{2017arXiv170208322A}.  Moreover, these magnetic fields can lead to surface abundance inhomogeneities (i.e., spots) by suppressing thermal convection at the stellar surface, as seen from tomographic images \citep[e.g., HD\,75049;][]{2015A+A...574A..79K}, the rotational modulation of absorption lines \citep[e.g., HR\,5907;][]{2012MNRAS.419.1610G} or from the photometric brightness \citep[e.g., $\sigma$\,Ori\,E;][]{2015MNRAS.451.2015O}.

In addition, the stellar wind is affected by the large-scale magnetic field, as ionized ejected mass follows the magnetic field lines, leading to typical observational signatures.  Depending on the strength of the magnetic field, its exact orientation, the mass-loss rate, and the rotation velocity, the wind particles can remain trapped in the circumstellar environment, that is, they form a centrifugal magnetosphere, or fall back on the star from a density enhanced region while being constantly renewed, that is, they form a dynamical magnetosphere \citep[e.g.,][]{2002ApJ...576..413U, 2005MNRAS.357..251T}.  There is often direct observational evidence for magnetospheres, such as, rotationally modulated X-ray emission, coming from shocks between the wind particles arriving at the magnetic equator from both hemispheres, or rotationally modulated H$\alpha$ emission from accumulated plasma.  Moreover, the large-scale magnetic field considerably amplifies the angular momentum loss through the stellar wind \citep[][]{1962AnAp...25...18S}.  This effect has been quite extensively modeled for the radiative wind of massive stars \citep[e.g][]{2002ApJ...576..413U, 2004ApJ...600.1004O}.

\begin{table*}[t]
\caption{Observing log of the Narval observations.  The exposure time of the full sequence, as well as the mid-exposure HJD, are indicated.  The rotation phase, $\phi_{\rm rot}$ is determined with $P_{\rm rot} = 0.897673$d and $T_0 = 2456185.8380$\,d.  The provided S/N is that of the LSD Stokes\,I profile calculated with various linemasks.  In addition, the magnetic detection status for each observation is indicated (DD = Definite Detection, MD = Marginal Detection, and ND = Non Detection).}
\centering
\tabcolsep=6pt
\begin{tabular}{cccccccccc}
\hline
\hline
ID	& HJD [d]	& $t_{\rm exp}$ [s]	& $\phi_{\rm rot}$	& \multicolumn{2}{c}{all}&	\multicolumn{2}{c}{He only}& \multicolumn{2}{c}{He excluded}\\
	& -2450000	&		&				  	& S/N	&Detect.& S/N	&Detect. & S/N	&Detect. \\
\hline
1	&	6185.66197234	&	4x1000s	&	0.803874	&	2067	&	DD	&	297	&	ND	&	1597	&	ND	\\
2	&	6203.61791407	&	4x1000s	&	0.806639	&	2075	&	DD	&	291	&	MD	&	1666	&	MD	\\
3	&	6206.64674915	&	4x1000s	&	0.180735	&	2111	&	DD	&	288	&	DD	&	1662	&	MD	\\
4	&	6214.58370424	&	4x1000s	&	0.022434	&	2032	&	ND	&	290	&	ND	&	1647	&	ND	\\
5	&	6230.56793210	&	4x800s	&	0.828727	&	2083	&	DD	&	288	&	DD	&	1650	&	MD	\\
6	&	6230.60771309	&	4x800s	&	0.873043	&	2052	&	DD	&	284	&	MD	&	1657	&	MD	\\
7	&	6230.64747616	&	4x800s	&	0.917338	&	2032	&	DD	&	280	&	MD	&	1654	&	MD	\\
8	&	6232.54916364	&	4x800s	&	0.035802	&	1972	&	MD	&	284	&	MD	&	1624	&	ND	\\
9	&	6232.58891699	&	4x800s	&	0.080087	&	2018	&	MD	&	284	&	ND	&	1634	&	MD	\\
10	&	6232.62864977	&	4x800s	&	0.124349	&	2012	&	DD	&	283	&	DD	&	1643	&	ND	\\
11$^a$	&	6244.56044152	&	4x800s	&	0.416262	&	1964	&	MD	&	286	&	MD	&	1590	&	ND	\\
12	&	6244.60021716	&	4x800s	&	0.460572	&	2018	&	MD	&	286	&	ND	&	1620	&	ND	\\
13	&	6244.64000450	&	4x800s	&	0.504894	&	2050	&	MD	&	288	&	ND	&	1611	&	ND	\\
14	&	6245.63552455	&	4x800s	&	0.613895	&	1734	&	ND	&	287	&	ND	&	1454	&	ND	\\
15	&	6245.67535221	&	4x800s	&	0.658263	&	1703	&	ND	&	288	&	ND	&	1412	&	ND	\\
16$^a$	&	6245.71511288	&	4x800s	&	0.702556	&	1753	&	ND	&	289	&	ND	&	1485	&	ND	\\
17	&	6254.51437756	&	4x800s	&	0.504861	&	1488	&	ND	&	275	&	ND	&	1277	&	MD	\\
18	&	6254.55412207	&	4x800s	&	0.549136	&	1959	&	MD	&	288	&	ND	&	1571	&	ND	\\
19	&	6254.59394714	&	4x800s	&	0.593501	&	1873	&	ND	&	287	&	ND	&	1510	&	ND	\\
20	&	7326.61890886	&	4x800s	&	0.820086	&	1991	&	MD	&	296	&	ND	&	1506	&	ND	\\
21	&	7326.66079329	&	4x800s	&	0.866745	&	1987	&	MD	&	294	&	ND	&	1515	&	MD	\\
22	&	7326.70220134	&	4x800s	&	0.912874	&	1906	&	ND	&	292	&	ND	&	1468	&	DD	\\
23	&	7337.61241188	&	4x800s	&	0.066754	&	1914	&	MD	&	276	&	MD	&	1458	&	MD	\\
24	&	7337.65343636	&	4x800s	&	0.112455	&	1911	&	DD	&	275	&	MD	&	1461	&	ND	\\
25	&	7337.70094841	&	4x800s	&	0.165383	&	1929	&	MD	&	281	&	MD	&	1488	&	ND	\\
26$^a$	&	7339.61360976	&	4x800s	&	0.296071	&	1938	&	DD	&	278	&	MD	&	1541	&	ND	\\
27$^a$	&	7339.65460177	&	4x800s	&	0.341736	&	1872	&	ND	&	273	&	ND	&	1482	&	ND	\\
28$^b$	&	7339.69683246	&	4x800s	&	0.388781	&	1401	&	ND	&	247	&	ND	&	1475	&	ND	\\
29	&	7412.46045929	&	4x800s	&	0.446835	&	1907	&	MD	&	281	&	ND	&	1513	&	ND	\\
30	&	7412.50162638	&	4x800s	&	0.492695	&	1934	&	ND	&	287	&	ND	&	1536	&	ND	\\
31	&	7412.54282594	&	4x800s	&	0.538591	&	1927	&	ND	&	292	&	ND	&	1513	&	MD	\\
32	&	7413.45749773	&	4x800s	&	0.557527	&	1923	&	MD	&	285	&	MD	&	1554	&	ND	\\
33	&	7413.49866839	&	4x800s	&	0.603391	&	1864	&	ND	&	290	&	ND	&	1511	&	ND	\\
34	&	7413.53974771	&	4x800s	&	0.649153	&	1942	&	MD	&	293	&	ND	&	1558	&	ND	\\
\hline
\end{tabular}
\label{tab:narval_log}
\tablefoot{${a}$: These observations were used in Sect.\,\ref{sec:longfield} together with the rotation phase-binned measurements from Table\,\ref{tab:phasebin_log} to fully cover the rotation period.  ${b}$: This observation was excluded during the analysis, since two of the sub-exposures were particularly noisy, leading to issues in the corresponding Stokes\,V profile with a very low S/N.}
\end{table*}

The most important consequence of a strong large-scale magnetic field, perhaps, is the altered internal stellar structure when the Lorentz force competes with the pressure force and gravity, directly influencing the exact evolution of the (massive) star.  Indeed, theoretical work indicates that the magnetic field acts upon the mixing processes in the radiative layers, by enforcing (quasi-)uniform rotation \citep[e.g.,][]{1937MNRAS..97..458F, 1992MNRAS.257..593M}, contradictory to what is expected from the enhanced angular momentum loss.  Because the radiative zone rotates uniformly and because of the fossil magnetic field itself, the physical processes in the layer around the boundary with the convective core are influenced.  The penetration depth of convective structures (i.e., plumes) with inertia \citep[e.g.,][]{1981ApJ...245..286P, 2004ApJ...601..512B} is expected to be smaller than in absence of a magnetic field, leading to a smaller convective-core-overshooting region.  

Observational evidence for the internal chemical and structural conditions is not easily available.  At present, only asteroseismology, through the interpretation of detected stellar pulsation modes, permits such a direct probing of the stellar interior.  Fortunately, some of the known magnetic early-type stars exhibit stellar pulsation modes and, thus, provide these important diagnostics.  An overview of the currently known magnetic early-type pulsators is given in \citet{2017magneticiaus_inprep}.  So far, only V2052\,Oph \citep{2012A+A...537A.148N, 2012MNRAS.424.2380H, 2012MNRAS.427..483B} and $\beta$\,Cep \citep{2000ApJ...531L.143S, 2013A+A...555A..46H} have been studied in detail using magneto-asteroseismology.  Respectively three and five pulsation modes have been detected for these two magnetic pulsators.  Pressure modes most sensitive to the conditions inside the radiative envelope are concerned.  The target of the present study is HD\,43317, one of the magnetic early-type stars with a rich frequency spectrum of gravity modes.

\subsection{The pulsating magnetic hot star HD\,43317}
HD\,43317 has a B3.5V spectral type and has no detected binary companion \citep[][hereafter P12]{2012A+A...542A..55P}.  Its chemical surface abundances agree with the solar abundances, but with some co-rotating He abundance spots at the stellar surface.  By studying a 150\,d CoRoT \citep[COnvection ROtation and planetary Transits;][]{2006cosp...36.3749B} lightcurve, P12 detected both pressure and gravity mode frequencies, making HD\,43317 a SPB/$\beta$\,Cep hybrid pulsator, with rotational modulation.  Several of the detected pulsation frequencies fall into one of two frequency series with a constant period spacing of either 6339\,s or 6380\,s.  The derived rotation period from the CoRoT photometry is $0.8969\pm0.0053$\,d.  Similar results were obtained by P12 from the detailed frequency analysis of selected spectral lines of simultaneous HARPS spectroscopy \citep{2003Msngr.114...20M}, spanning approximately 25\,d.  High-degree $\beta$\,Cep pulsation modes were detected in the selected \ion{Mg}{II}\,4481$\AA$ line, while the \ion{He}{I} lines dominantly vary with twice the determined rotation frequency indicative of rotational modulation.

Since many of the extracted pulsation-mode frequencies from the CoRoT lightcurve fall below twice the deduced rotation frequency, these pulsation modes are unstable coherent internal gravito-inertial waves \citep[IGWs; ][and references therein]{2014A+A...565A..47M} possibly excited by the $\kappa$-mechanism, but perturbed by the Coriolis force, instead of regular gravity modes in a non-rotating non-magnetic star.  These IGWs are highly efficient at transporting angular momentum to the stellar surface \citep[e.g.,][]{2005MNRAS.364.1135R, 2013ApJ...772...21R}.  On the other hand, convection-driven IGWs are linked to the stochastic motion by the convective plumes escaping the convective core \citep[e.g.,][]{1990ApJ...363..694G, 2010Ap+SS.328..253S, 2013MNRAS.430.1736S, 2014A+A...565A..47M}.  Such stochastic IGWs have only recently been observed for massive stars \citep[e.g][]{2011A+A...533A...4B, 2012A+A...546A..47N, 2017A+A...602A..32A}.  In addition, it has become clear from \textit{Kepler} data that gravito-inertial modes in B- and F-type pulsators with period spacings are dipole modes \citep{2015A+A...574A..17V, 2017A+A...598A..74P}.  Only detailed modeling will be able to corroborate whether some of the detected stellar pulsation-mode frequencies of HD\,43317 are IGWs and hint towards their driving mechanism.

\citet{2013A+A...559A..25S} attempted to explain the detected pulsation-mode frequencies of HD\,43317 detected with CoRoT as unstable oscillations in a rapidly rotating star using numerical simulations.  Many of the high-visibility axisymmetric gravity modes fall in the correct frequency regime and nearly coincide with the observations by P12.  However, these unstable modes are not able to explain the observed period spacings, making it difficult to corroborate the assumption of unstable oscillations.  Moreover, during the analysis of \citet{2013A+A...559A..25S}, HD\,43317 was not yet known to host a large-scale magnetic field.

Following the detection of rotational modulation, especially for the He absorption lines, and the strong X-ray flux \citep{1996A+AS..118..481B}, a spectropolarimetric monitoring campaign for HD\,43317 was performed to look for the possible presence of a polarization signature due to a magnetic field.  This anticipated magnetic field was detected by \citet[][hereafter B13]{2013A+A...557L..16B}.  Enforcing the rotation period found in the CoRoT lightcurve by P12 on the variation of the longitudinal field measurements provided a good sinusoidal fit and indicated a dipolar configuration for the magnetic field.  The estimated field strength at the magnetic poles was about 1\,kG.  Less strong constraints were obtained on the orientation of the field.  By using the inclination angle to the stellar rotation axis determined by P12, namely $i \in [20, 50]\,^{\circ}$, B13 obtained the obliquity angle of the magnetic field to the rotation axis $\beta \in [70, 86]\,^{\circ}$.  These weak constraints are in part due to the limited rotation phase coverage and the long duration of the nightly averaged exposures of B13 with respect to the rotation phase.  Finally, the authors did not find any rotational modulation of emission lines indicating the presence of a magnetosphere, but the determined value for the magnetic confinement parameter $\eta_*$ implies that HD\,43317 has a centrifugal magnetosphere.

We have taken additional spectropolarimetric observations of HD\,43317 to increase the coverage over the rotation phase.  The goal is to improve the observational constraints on the strength and orientation of the large-scale magnetic field.  This updated characterization of the field will be used to perform detailed (magneto-)asteroseismic modeling in a following paper.

The obtained data, the studied observations, and the treatment thereof are discussed in Sect.\,\ref{sec:observations}.  First, we investigated how strongly the stellar pulsation modes and the rotational modulation influenced the Narval spectropolarimetry, and what the possible implications are (Sect.\,\ref{sec:LPV}).  In addition, we use the rotational modulation to update the rotation period of HD\,43317.  Sections\,\ref{sec:magnetic} and \ref{sec:vmodelling} are dedicated to the characterization of the large-scale magnetic field of HD\,43317, by analyzing the longitudinal magnetic field and modeling the Zeeman signatures with synthetic profiles of an oblique dipolar field, respectively.  The obtained results and their implications are discussed in Sect.\,\ref{sec:discussion} and conclusions are summarized in Sect.\,\ref{sec:conclusions}.

\begin{figure}[t]
		\centering
			\includegraphics[width=0.48\textwidth, height = 0.66\textheight]{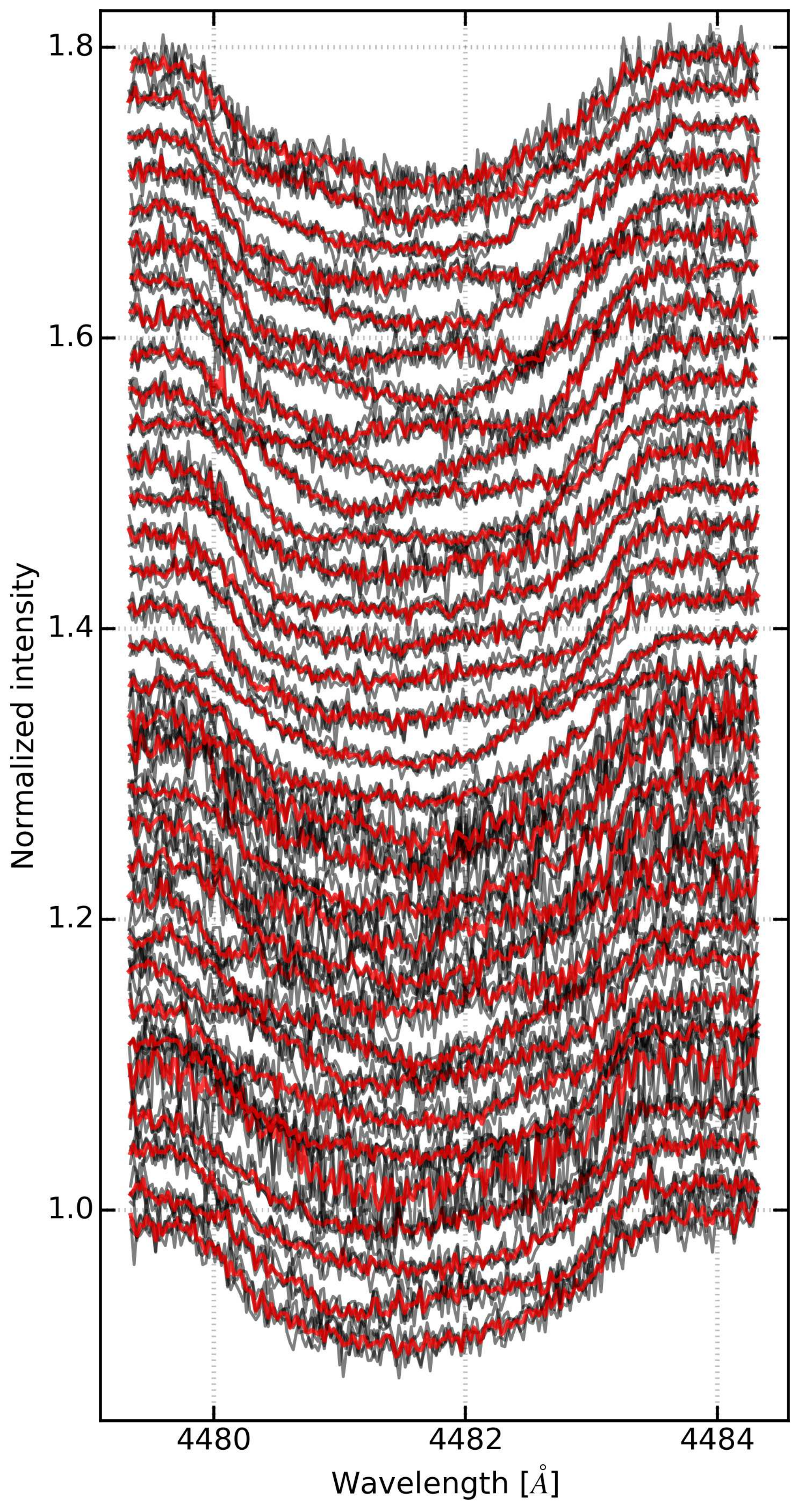}%
			\caption{Profiles of the \ion{Mg}{II}\,4481$\AA$ line for each Narval spectropolarimetric sequence in red (for which the four individual subexposures are given in black).  The profiles are organized  according to rotation phase, with a constant offset (independent of the phase difference) for visibility.}
			\label{fig:Mg4481}
\end{figure}
\begin{figure}[t]		
		\centering
			\includegraphics[width=0.48\textwidth, height = 0.66\textheight]{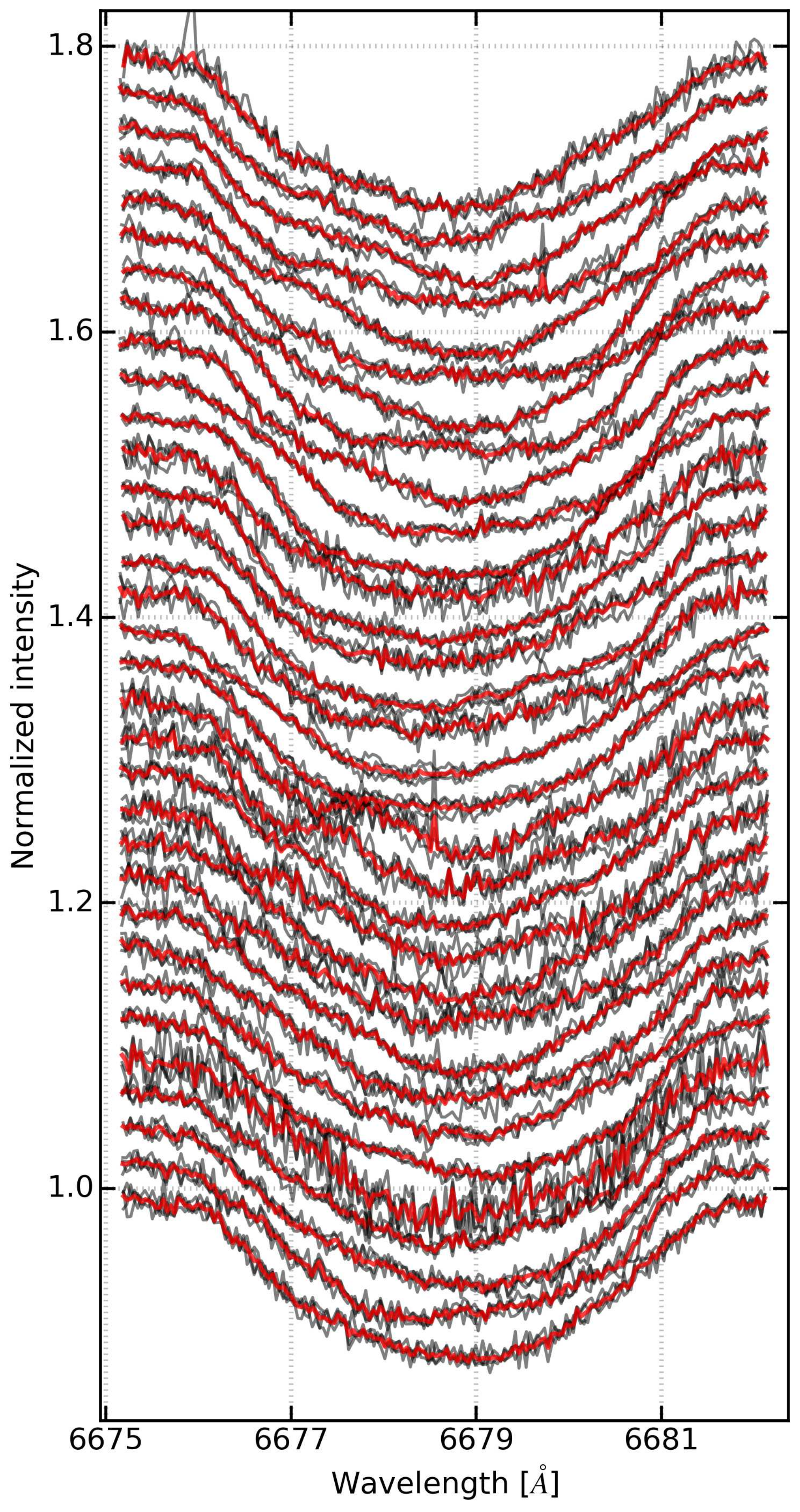}%
			\caption{As in Fig.\,\ref{fig:Mg4481}, but for the \ion{He}{I}\,6678$\AA$ line.}
			\label{fig:He6678}
\end{figure}

\section{Observations}
\label{sec:observations}
\subsection{Spectropolarimetry}
To detect the Zeeman signature of the magnetic field of HD\,43317, we have obtained spectropolarimetric observations with Narval \citep{2003EAS.....9..105A} mounted at T{\'e}lescope Bernard Lyot (TBL) at Pic du Midi in France.  These observations measure the circular polarization (Stokes\,V), and span from 3700\,$\AA$ to 10500\,$\AA$ with an average resolution of $\sim\,65000$.  Standard settings were employed, with bias, flat-field, and ThAr calibration images taken at both the beginning and end of each night.  We combined these new data, taken between November\,2015 and January\,2016, with the publicly available Narval spectropolarimetry discussed in B13, obtained with the same settings between 14 September 2012 and 22 November 2012.  Some of the earlier observations of B13 are single sequences of four sub-exposures with an exposure time of 1000\,s per sub-exposure.  Otherwise, the obtained spectropolarimetry consists of three consecutive sequences of four sub-exposures, each with an 800\,s integration time.  The full duration of an individual spectropolarimetric sequence was carefully tailored to avoid strong line-profile variations (LPVs) between the consecutive sub-exposures.   In total 34 sequences are available for HD\,43317, with an average signal-to-noise (S/N) between 100 and 325.  The complete overview of the data is given in Table\,\ref{tab:narval_log}.

All Narval data were reduced with the \textsc{libre-esprit} software \citep{1997MNRAS.291..658D}, available at the TBL.  A spectropolarimetric observation was deduced per sequence, as well as a spectroscopic measurement for each of the individual Narval sub-exposures.  The spectropolarimetric data were rectified to unity continuum using interactive spline fitting tools with IRAF\footnote{IRAF is distributed by the National Optical Astronomy Observatory,which is operated by the Association of Universities for Research in Astronomy (AURA) under a cooperative agreement with the National Science Foundation.}.  The spectroscopic observations were rectified and normalized per spectral order as in P12.

The obtained Narval spectropolarimetry forms the basis for the characterization of the large-scale magnetic field at the stellar surface of HD\,43317 (Sect.\,\ref{sec:magnetic}).

\subsection{Spectroscopy}
We also downloaded the publicly available and fully reduced HARPS spectroscopy of HD\,43317, taken in the 'EGGS' configuration with the ESO 3.6-m telescope in La Silla in Chile.  These observations cover the wavelength range 3780 -- 6910\,$\AA$ with a resolving power of about 80000.  The exposure times of the HARPS spectroscopy vary between 240 and 600\,s, leading to an average S/N between 150 and 300.  In total, 191 measurements were obtained, taken over a time span of 25\,days in December\,2009.  These observations were rectified and normalized to unity continuum using the same tools as for the Narval spectroscopy.  However, this time, the normalization process was performed on the complete wavelength range, because the consecutive spectral orders were accurately merged.

The Narval and HARPS spectroscopic datasets were employed to study the LPVs of several spectral lines in Sect.\,\ref{sec:LPV}.

\section{Line profile analysis}
\label{sec:LPV}
\subsection{Line profile variability}
\label{sec:LPV_pol}
As discussed by P12, HD\,43317 shows obvious periodic LPVs.  These variations are in part due to the stellar pulsation modes and in part due to the rotational modulation through co-rotating surface abundance inhomogeneities.  We selected both the \ion{Mg}{II}\,4481$\AA$ and the \ion{He}{I}\,6678$\AA$ lines, similarly to P12, to investigate the stellar pulsation modes and the rotational modulation, respectively.

Figure\,\ref{fig:Mg4481} shows the individual and average line profiles for each sequence for the \ion{Mg}{II}\,4481$\AA$ line, organized by rotation phase (determined in Sect.\,\ref{sec:Prot}).  From this visualization, we note that a few sub-exposures exhibit mild variations during a given polarimetric sequence.  However, for almost all observations, these effects of the LPVs are comparable to the noise level, with intensity changes only up to $1\,\%$.

Comparing the mean profiles for various sequences clearly shows the variability of the star.  One sequence taken on November 12, 2015 shows a poor S/N.  We have excluded this sequence for the remainder of the analyses (and Figs.\,\ref{fig:Mg4481} and \ref{fig:He6678}).  Similar results on the extent of the LPVs during each sub-exposure were obtained when comparing the mean profiles for the \ion{He}{I}\,6678$\AA$ line, given in Fig.\,\ref{fig:He6678}, which is organized in a similar manner as Fig.\,\ref{fig:Mg4481}.

\begin{figure}[t]
		\centering
			\includegraphics[width=0.48\textwidth, height = 0.33\textheight]{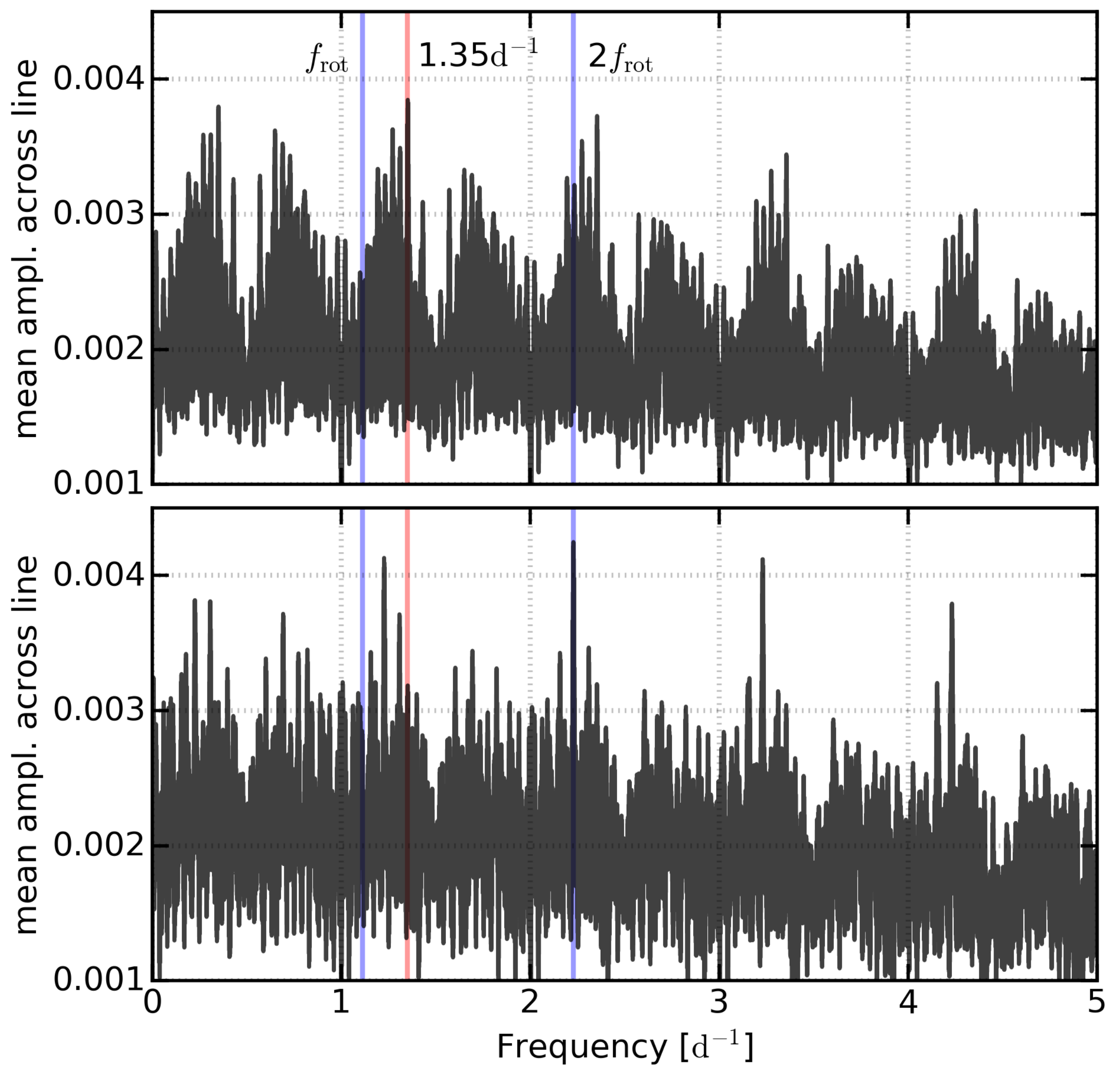}%
			\caption{Mean discrete Fourier periodograms over the full line profile variations for the \ion{Mg}{II}\,4481$\AA$ (\textit{top}) and the \ion{He}{I}\,6678$\AA$ (\textit{bottom}) absorption lines in the Narval spectroscopy, determined with the pixel-by-pixel method.  The rotation frequency and the second harmonic of the rotation frequency are indicated in blue.  The found pulsation-mode frequency in the \ion{Mg}{II} line is indicated in red.  Strong 1\,\d aliasing frequency peaks are present in the periodograms.}
			\label{fig:periodogram_LPV}
\end{figure}
\begin{figure}[t]
		\centering
			\includegraphics[width=0.48\textwidth, height = 0.33\textheight]{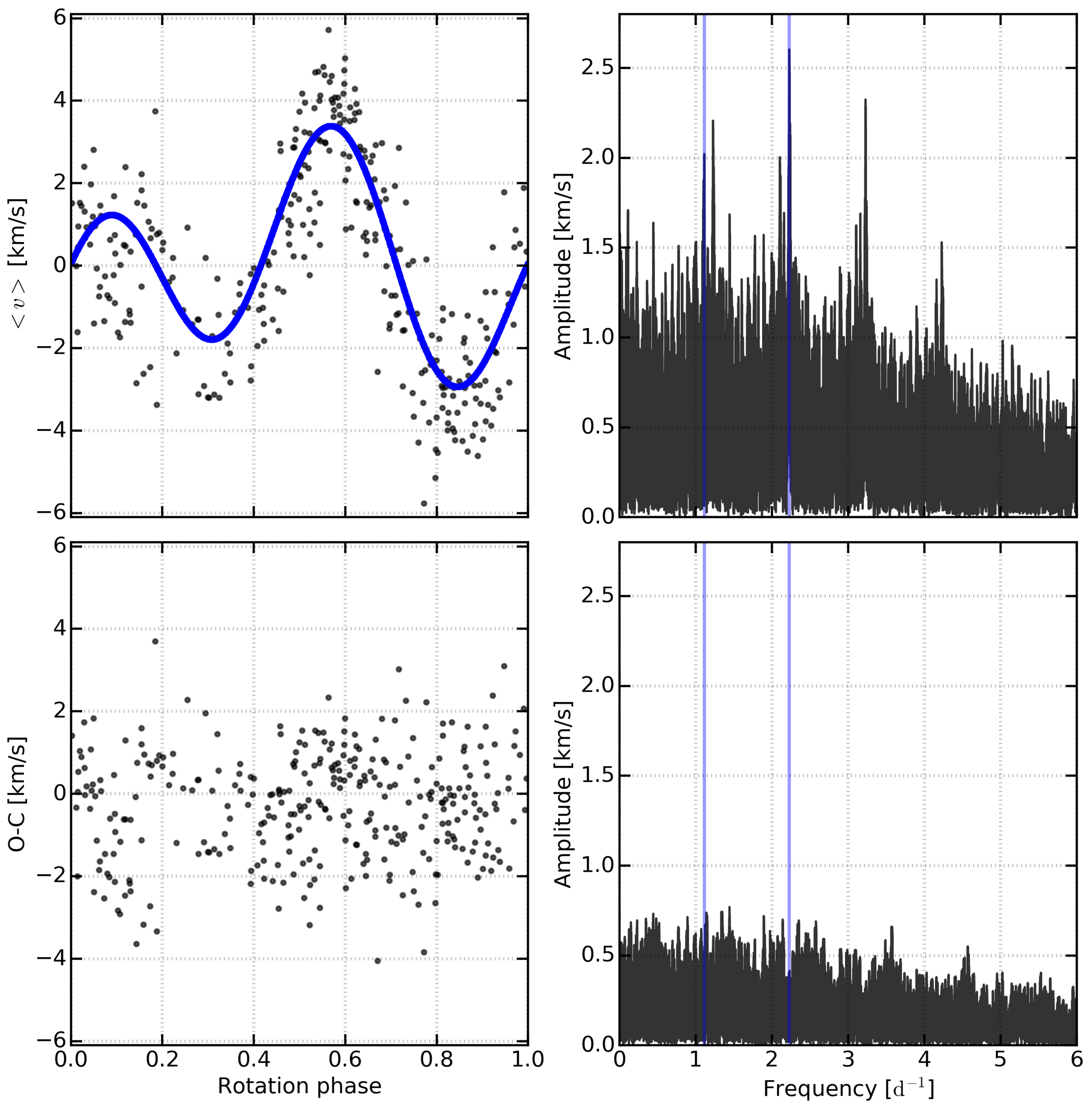}%
			\caption{The first moment $<v>$ of the \ion{He}{I}\,6678$\AA$ line from the combined Narval and HARPS spectroscopy (\textit{top left}).  The model for the rotational modulation (see Eq.\,(\ref{eq:secondordersine})) for  $P_{\rm rot} = 0.897673$\,d and $T_0 = 2456185.8380$\,d is given as a solid blue line and the observations as black dots.  \textit{Bottom left}: residuals of $<v>$ after the subtraction of the model .  \textit{Right}: Computed Lomb-Scargle periodograms of $<v>$ (\textit{top}) and the residuals (\textit{bottom}), with $f_{\rm rot}$ and $2f_{\rm rot}$ marked as vertical blue lines.}
			\label{fig:firstmomement}
\end{figure}

Because the stellar pulsation periods and the rotation period are expected to remain constant over the time span of a decade, we performed a detailed frequency analysis on the LPVs of the full Narval dataset.  For this analysis, we used the \textsc{famias} software package \citep{2008CoAst.155...17Z} and determined the discrete Fourier periodogram for the pixel-by-pixel method \citep{2000ASPC..210..138M}.  \textsc{famias} computes the discrete Fourier periodogram at each velocity step within the absorption line and averages this to a mean periodogram capturing the periodic LPVs.  These discrete Fourier periodograms are shown in Fig.\,\ref{fig:periodogram_LPV} for both the \ion{Mg}{II}\,4481$\AA$ and \ion{He}{I}\,6678$\AA$ lines.  For the \ion{Mg}{II}\,4481$\AA$ line, we retrieved variability with the known pulsation mode at $f=1.35$\,\d from P12.  However, the amplitude of the frequency did not pass the S/N threshold of 4, calculated over a 1\,\d frequency domain, following the typical S/N criterion for significance used during detailed frequency extractions \citep{1993A+A...271..482B}.  For the \ion{He}{I}\,6678$\AA$ line, we recovered the known rotational modulation with $2f_{\mathrm{rot}}\approx2.23$\,\d, which was determined to be significant following the S/N criterion.  Adopting a prewhitening approach, we removed the above mentioned variability using the pixel-by-pixel method.  Subsequent frequency extractions did not indicate clear results, most likely because of the poor temporal coverage of the Narval data with respect to the stellar pulsation-mode frequencies.

\subsection{Determining the rotation period}
\label{sec:Prot}
P12 have indicated that the abundance inhomogeneities at the stellar surface of HD\,43317 cause rotational modulation of certain spectral lines, especially the \ion{He}{} lines.  In the previous sub-section, we have shown that this is also observed in the Narval spectroscopy.  We used these LPVs to constrain the rotation period with great accuracy and precision, under the assumptions that only two surface inhomogeneities are present, typical for magnetic early-type stars due to a dominant dipolar magnetic field, that these features remain constant on a timescale of a decade and that they do not migrate over the stellar surface.  By combining the HARPS and Narval spectroscopy, we obtained a dataset spanning 6.13\,years, with 325 measurements, distributed between three observing campaigns.

Employing \textsc{famias}, we isolated the \ion{He}{I}\,6678$\AA$, which exhibits clear rotational modulation, and deduced the first moment of the line, $<v>$.  When phase-folding $<v>$ with the rotation period, the modulation should follow a second-order sine:
\begin{equation}
\Magr(t) =  A_1 \sin (2\pi(f_{\rm rot}t + \phi_1)) + A_2\sin(2\pi(2f_{\rm rot}t + \phi_2)) + C\,\mathrm{,}
\label{eq:secondordersine}
\end{equation}
\noindent with $A_i$ and $\phi_i$ being the amplitudes and phases of the individual sine terms, $C$ a constant offset, and $f_{\rm rot}$ the rotation frequency.  As an initial estimate, we performed a Least-Squares (LS) minimization fit to the observed $<v>$ of the \ion{He}{I}\,6678$\AA$ line of HD\,43317.  In a subsequent step, a Bayesian Monte Carlo Markov Chain (MCMC) approach \citep[using \textsc{emcee};][]{2013PASP..125..306F} was adopted to obtain a more reliable result for the rotation period.  Uniform priors were considered in the appropriate parameter spaces, and we adjusted the log-likelihood from \citet{1986ssds.proc..105D} and \citet{1990ApJ...364..699A} to include individual weights:
\begin{equation}
\Lagr(\Theta) = \sum\limits_{i=1}^{N}\left\{ w_i \left(\ln \left(|\Magr(\Theta; t_i)|\right) + \frac{|<v>(t_i)|}{|\Magr(\Theta; t_i)|}\right)\right\} \ \mathrm{,}
\label{eq:loglikelihood}
\end{equation}
\noindent with $\ln$ being the natural logarithm, $\Magr(\Theta; t_i)$ the model of Eq.\,(\ref{eq:secondordersine}) for a given parameter vector $\Theta$, $<v>(t_i)$ the observed first moment at $t_i$, and $w_i$ the weight for that particular observation deduced from the S/N of the corresponding spectroscopic observation.  Calculations with the Bayesian MCMC were started in a Gaussian ball around the LS result using 500 parameter chains and continued until the full parameter space was sampled and stable solutions were obtained.

The posterior probability distribution for the rotation period indicated $P_{\rm rot} = 0.897673 \pm 0.000004$\,d as the most likely period causing the rotational modulation of $<v>$ of the \ion{He}{I}\,6678$\AA$ line.  No clear posterior probability distributions were obtained for the remaining model parameters.  Hence, we accepted the rotation period from the MCMC and updated the remaining parameters in Eq.\,(\ref{eq:secondordersine}) by means of a LS minimization fit to $<v>$.  The final model for the variability of the measured $<v>$ is shown in Fig.\,\ref{fig:firstmomement}, as well as the residuals to the fit.  Some scatter remains for the residuals, yet the Lomb-Scargle periodogram \citep{1976Ap+SS..39..447L, 1982ApJ...263..835S} of the residuals indicates no remaining periodic variability.  The adopted value for $P_{\rm rot}$ agrees well with the literature rotation period determined from the CoRoT photometry (P12).  Enforcing this literature value for $P_{\rm rot}$ on $<v>$, however, resulted in a worse LS minimization model for $<v>$ than our derived solution.  The zeroth moment of the absorption lines, that is, the equivalent width, and the second moment, the line width, were also investigated for periodic variability related to the stellar rotation.  However, these two diagnostics show a much larger scatter compared to the first moment and no strong amplitude peaks in their respective periodograms.  Thus, we were unable to accurately determine $P_{\rm rot}$ from either $<v^0>$ or $<v^2>$, possibly due to the stellar pulsation modes.

\begin{figure}[t]
		\centering
			\includegraphics[width=0.48\textwidth, height = 0.66\textheight]{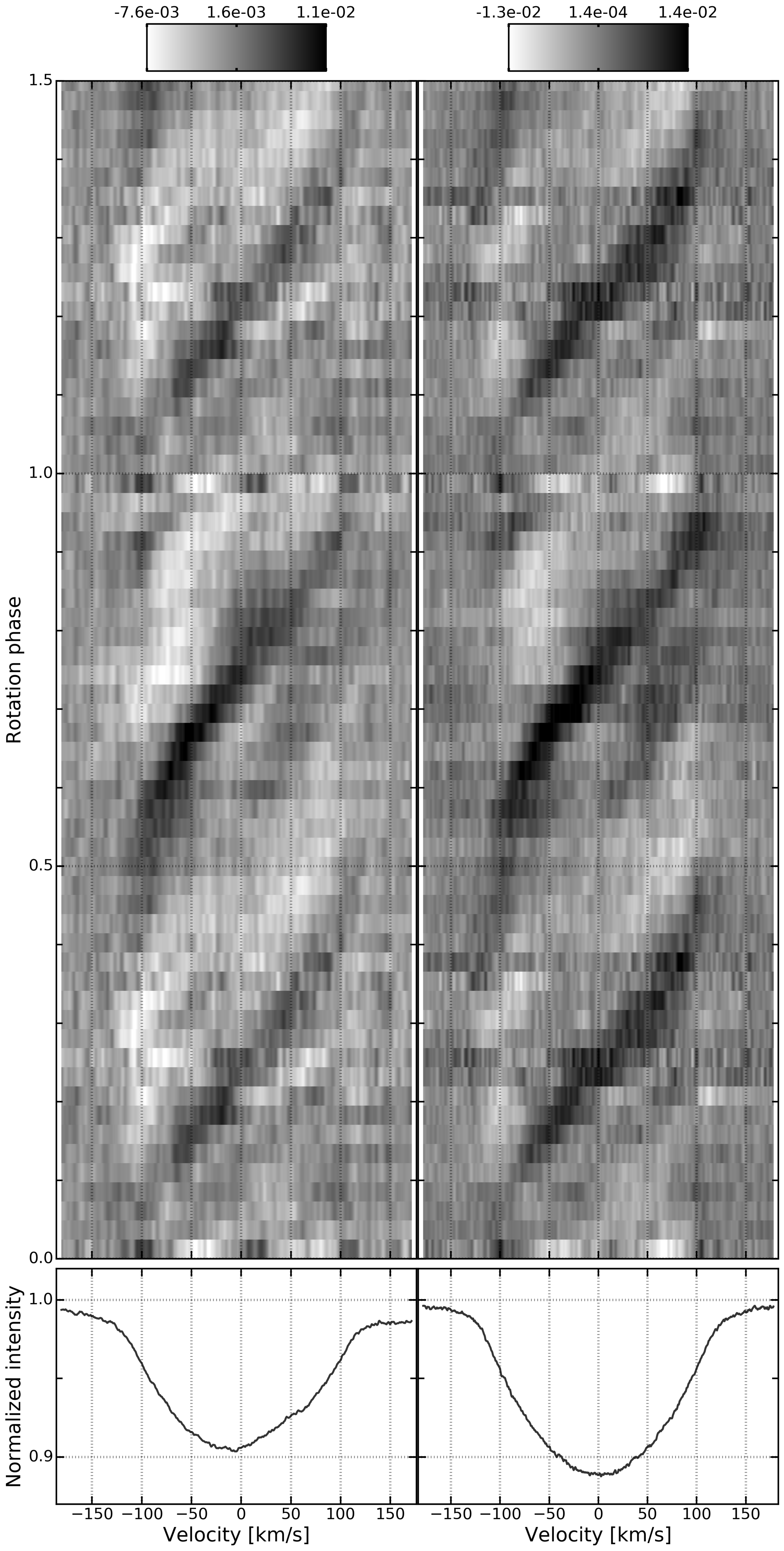}%
			\caption{\textit{Top}: Dynamical spectra of the residual \ion{He}{I}\,5016$\AA$ (\textit{left}) and \ion{He}{I}\,6678$\AA$ (\textit{right}) lines to their respective median profile, phase folded with the rotation period $P_{\rm rot} = 0.897673$\,d and $T_0 = 2456185.8380$\,d, and smoothed slightly with a Gaussian filter for increased visibility.  The variability corresponds to at least two co-rotating surface inhomogeneities at the stellar surface, causing the rotational modulation.  The amplitude of the residuals to the mean profile is indicated by the gray-scale.  \textit{Bottom}: Mean line profile of both \ion{He}{I} absorption lines.}
			\label{fig:dynamicalHe}
\end{figure}

With the rotation period now firmly established, we investigate the co-rotating surface inhomogeneities themselves.  In particular, we wish to provide some constraints on their position on the stellar surface.  Therefore, we constructed a dynamical spectrum with both the Narval and HARPS spectroscopy of the \ion{He}{I}\,6678$\AA$ line, indicating the residual absorption line to the mean line profile at a given rotation phase (right panel of Fig.\,\ref{fig:dynamicalHe}).  We considered 50 rotation phase bins and constructed the residuals from a weighted average spectrum for all observations falling within that phase bin.  Two obvious features travel in velocity and rotation phase through the line profile, typical for co-rotating surface inhomogeneities.  Since the two features are separated about 0.5 in rotation phase, they are expected to reside at opposite sides of the star.  Both features show a slightly different duration with respect to the rotation phase, indicating that they are not located at the rotation equator, irrespective of the inclination angle to the rotation axis.  Moreover, the surface inhomogeneities are not located too close to the (rotation) poles, otherwise only one feature would have been visible (except when $i\approx 90\,^{\circ}$).  As such, we anticipate that the co-rotating He patches are located at a considerable obliquity angle $\beta_{\rm spot}$ with respect to the rotation axis, most likely $\beta_{\rm spot} \in [60, 80]^{\circ}$.

Lastly, two additional features are present at the rotation phase of $\sim\,0.7$ with positive velocity and at the rotation phase of $\sim\,0.95$ with negative velocity (marked as dark gray in Fig.\,\ref{fig:dynamicalHe}).  Since it is not clear whether these travel over the full spectral line, their origin is difficult to identify.  Detailed spot modeling, while accounting for the stellar pulsation modes, or tomographic imaging could aid in the characterization of this feature, but requires higher S/N spectroscopy covering the pulsational beating cycle.

We obtain similar conclusions when studying the rotational modulation of the \ion{He}{I}\,5016$\AA$ line (left panel of Fig.\,\ref{fig:dynamicalHe}).

\section{Magnetic field measurements}
\label{sec:magnetic}
\subsection{Zeeman signature}
\label{sec:zeeman}

\begin{figure*}[t]
		\centering
			\includegraphics[width=\textwidth, height = 0.33\textheight]{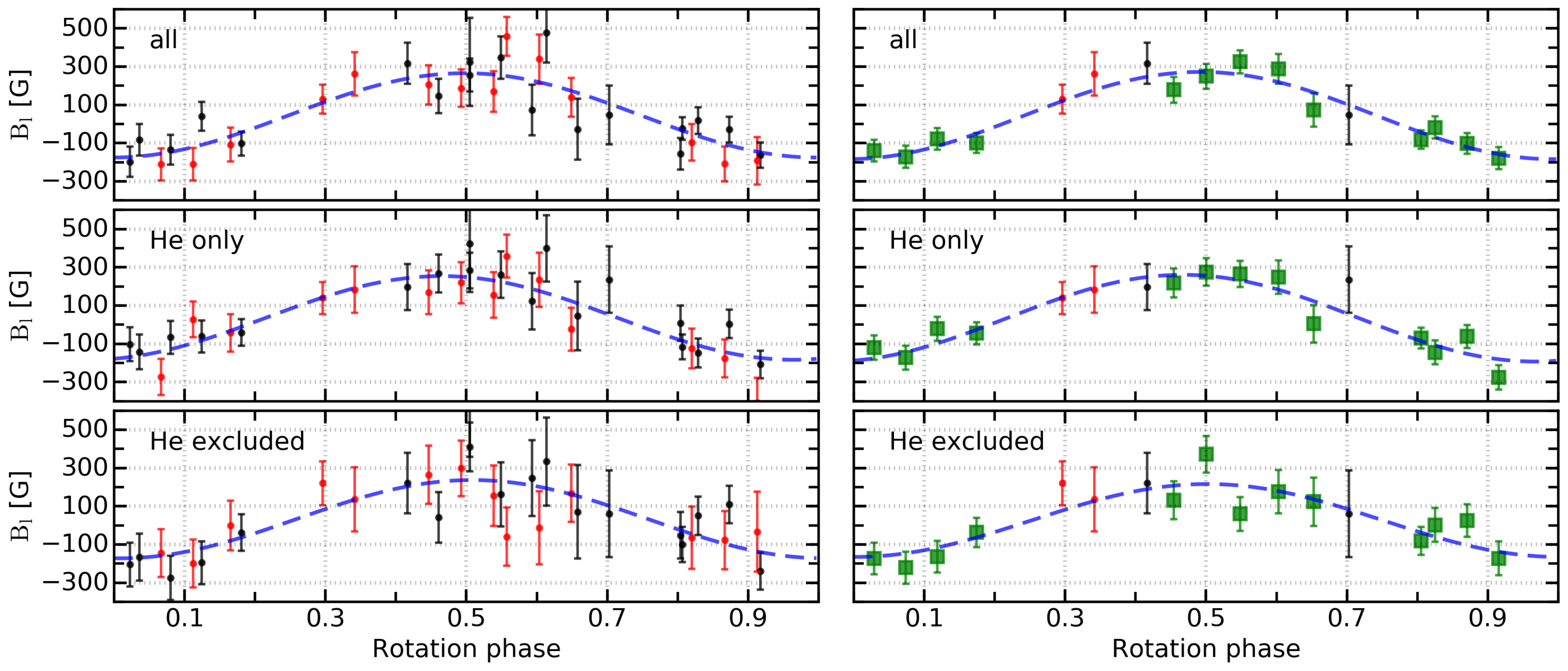}%
			\caption{Longitudinal field measurements for HD\,43317 for various LSD linemasks.  \textit{Left}: Deduced values from the individual Narval spectropolarimetric sequences.  The archival spectropolarimetry (B13) is indicated in black, while the additional data presented here are indicated in red.  The optimal LS minimization fit, representing the rotational modulation of a dipolar field, is given by the dashed blue line.  We assumed $P_{\rm rot} = 0.897673$\,d and $T_0 = 2456185.8380$\,d. \textit{Right}:  Using the phase-binned data, indicated by the green squares.}
			\label{fig:longmag}
\end{figure*}

To study the Zeeman signature of the magnetic field of HD\,43317, we constructed a mean line profile, employing the Least-Squares Deconvolution (LSD) technique \citep{1997MNRAS.291..658D}.  As such, the S/N of the polarization signature is greatly improved.  To perform the LSD method, we started from a VALD3 linemask \citep{2015PhyS...90e4005R} for a star of 17000\,K and a $\log g$ = 4.0\,dex, compatible with the literature values for HD\,43317 (P12).  Hydrogen lines as well as all metal lines blended with hydrogen lines, strong telluric features, and diffuse interstellar bands were removed from the linemask.  Lastly, the depths of the lines in the mask were adjusted to match the observations.  In addition, any lines with a depth smaller than 0.1 were discarded.  The final linemask includes 449 lines, with their wavelength, depth, and Land{\'e} factor, used in the LSD method.  We further subdivided the linemask into a version excluding all (blends with) He lines (containing 396 lines) and a version only containing the 17 He lines.

For each observation, the False Alarm Probability (FAP) was computed \citep{1997MNRAS.291..658D}, indicating the detection probability of a Zeeman signature in the LSD Stokes\,V spectrum.  Definitive detections (DD) show a clear signal and have a FAP $ < 10^{-3}\,\%$, marginal detections (MD) have $10^{-3}\,\% < \mathrm{FAP} < 10^{-1}\,\%$, while non-detections (ND) have a FAP $>10^{-1}\,\%$.  Using the full linemask, 9 observations are DD, 12 are MD, and 13 are MD (Table\,\ref{tab:narval_log}).  Such a low number of (clear) detections was expected, since the exposure times were tailored to avoid influence of the stellar variability, instead of reaching a high S/N.

As indicated in Sect.\,\ref{sec:LPV_pol}, a few of the spectropolarimetric observations are distorted by the LPVs.  Since the effect of stellar pulsation and rotational modulation is slightly different for each spectral line, the LSD method averages out any strong variability.  The diagnostic Null profile, which is constructed by destructively co-adding the consecutive sub-exposures of a polarimetric sequence, can indicate the effects of LPVs between these sub-exposures.  These Null profiles show some scatter, yet the level of the scatter is similar to what is found outside the lines (see Appendix\,\ref{sec:appendix_LSD}).   The amount of scatter is rather large, because of the limited S/N of the observations, but no obvious evidence of strong distortions by the LPVs during a spectropolarimetic sequence was noted.  These results agree with the conclusions from the LPVs in the individual exposures (we refer to Sect.\,\ref{sec:LPV_pol}).

\subsection{Longitudinal field measurements}
\label{sec:longfield}
The large-scale fossil magnetic fields of early-type stars are known to be stable over a timescale of at least a decade.  As the star rotates, the line-of-sight component of these magnetic fields, integrated over the stellar surface, varies with the rotation phase.  This longitudinal magnetic field \citep{1979A+A....74....1R} is measured as:

\begin{figure*}[p!]
		\centering
			\includegraphics[width=\textwidth, height = 0.90\textheight]{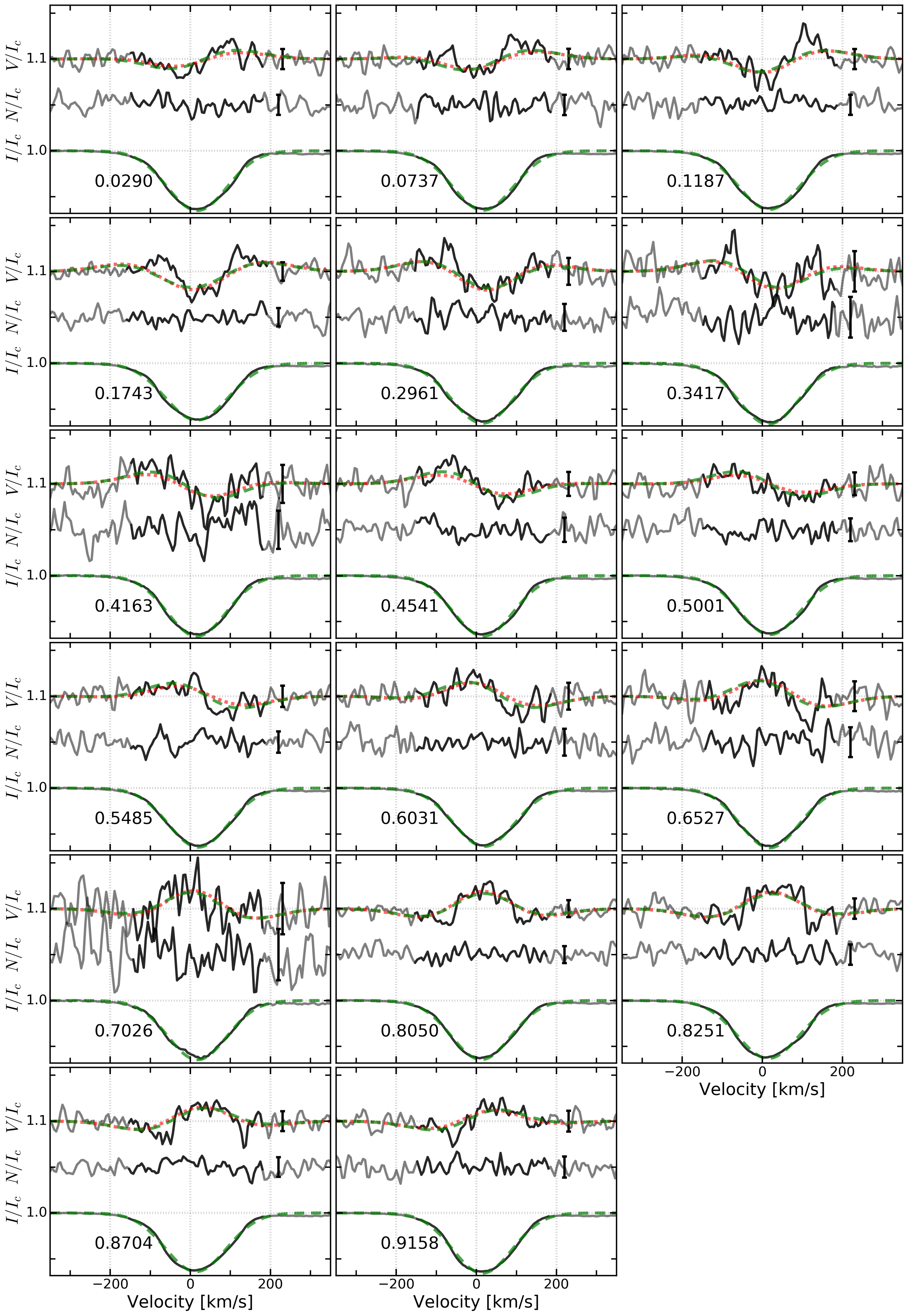}%
			\caption{Observed LSD profiles compared with the results from the Stokes\,V modeling.  Panels are organized according to the rotation phase and contain the observed intensity profile ($I/I_c$; \textit{bottom}), the Null profile ($N/I_c$; \textit{middle}), and the Stokes\,V ($V/I_c$; \textit{top}) profiles given in black.  Both the Null and Stokes\,V profiles are multiplied by a factor 100 and smoothed with a Gaussian filter with a width of 6\,km/s for increased visibility.  Their respective error bars are indicated on the right side of each panel.  The centered dipole model of Sect.\,\ref{sec:freevmodel} is indicated by the green dashed lines, while that of Sect.\,\ref{sec:restrictedvmodel} by the red dotted lines.}
			\label{fig:stokesVmodelling}
\end{figure*}

\begin{table}[t]
\caption{Obtained parameters from the dipolar fits to the longitudinal field measurements for HD\,43317.  The estimated obliquity angle, $\beta$, is provided, assuming $i \in [20, 50]^{\circ}$ from P12.  Employing these values, we deduced the limits on the strength of the dipolar magnetic field $B_{\rm dip}$.}
\centering
\tabcolsep=4pt
\begin{tabular}{lcccc}
\hline
\hline
mask& $B_0$	& $B_{\rm amp}$	& $\beta$	 & $B_{\rm dip}$	\\
	& [G]	& [G]			& [deg]		 & [G]\\
\hline
all				&$	45	\pm	16	$&$	221	\pm	22	$&$[	69	,	88	]$&$[	1013	,	2300	]$\\
He only			&$	35	\pm	18	$&$	219	\pm	24	$&$[	70	,	90	]$&$[	1003	,	2312	]$\\
He excl.			&$	33	\pm	25	$&$	205	\pm	34	$&$[	67	,	87	]$&$[	723	,	2150	]$\\
bin: all			&$	44	\pm	17	$&$	228	\pm	23	$&$[	69	,	89	]$&$[	1051	,	2403	]$\\
bin: He only		&$	34	\pm	19	$&$	227	\pm	25	$&$[	71	,	90	]$&$[	1045	,	2373	]$\\
bin: He excl.	&$	26	\pm	24	$&$	191	\pm	32	$&$[	68	,	88	]$&$[	657	,	1960	]$\\
\hline
\end{tabular}
\label{tab:longmag}
\end{table}

\begin{equation}
B_l = -2.14 \cdot 10^{11} \mathrm{G} \frac{\int v V(v) \mathrm{d}v}{\lambda g c \int[1-I(v)] \mathrm{d}v} \, \mathrm{,}
\label{eq:longfield}
\end{equation}
\noindent where $V(v)$ and $I(v)$ are the LSD Stokes\,V and I profiles for a given velocity $v$, $g$ is the mean Land{\'e} factor, $\lambda$ the mean wavelength (in nm) from the LSD method, and $c$ the speed of light.  The mean Land{\'e} factor is 1.2072, 1.1567, and 1.2095 for the LSD profiles constructed with the complete linemask, the He-only linemask, and the He-excluded linemask, respectively.  The corresponding mean wavelengths are 506.98\,nm, 494.04\,nm, and 511.58\,nm, respectively.  The integration limits should span the full Stokes\,I profile.  Overly large integration limits, however, will artificially increase the uncertainty on the measured $B_l$.  We opted for an integration limit of 200\,km/s, 115\,km/s, and 240\,km/s around the line center for the LSD profiles with the full linemask, He-excluded, and He-only linemasks, respectively.  These limits differ from the literature $v\sin i = 130 \pm 16$\,km/s \citep{2005ESASP.560..571G} or $v\sin i = 115 \pm 9$\,km/s (P12) whenever He lines are included.  These larger integration limits for the determination of the longitudinal magnetic field were needed to fully cover the LSD Stokes profiles.  This additional non-rotational broadening is most likely due to the pressure broadening of the He lines through the Stark effect \citep[e.g.,][]{1984A+A...136..289D}.

For a centered dipolar magnetic field, the rotationally modulated longitudinal field follows a sinusoidal pattern.  We performed an LS minimization fit to the measured $B_l$ values assuming $P_{\rm rot} = 0.897673 \pm 0.000004$\,d and $T_0 = 2456185.8380 \pm 0.0059$\,d ($T_0$ was determined from the subsequent Stokes\,V modeling in Sect.\,\ref{sec:freevmodel}).  These models for the longitudinal magnetic field are given in the left panel of Fig.\,\ref{fig:longmag}.  Despite the uncertainties on the $B_l$ values, we obtain a relatively good description of the modulation thanks to the sufficient phase coverage. In addition, we investigated the possibility of a quadrupolar contribution to the large-scale magnetic field by fitting a second order sine function (Eq.\,(\ref{eq:secondordersine})) to the measured $B_l$.  These fits all indicate that a quadrupolar contribution is negligible with the accepted rotation period.  The values for the amplitude of the dipolar model are given in Table\,\ref{tab:longmag}.

To decrease the uncertainty on the $B_l$ measurements, we phase-binned the Narval data according to their rotation phase.  When multiple observations differed by less than 2\,\% in rotation phase, we averaged their spectropolarimetric observations, using the S/N determined in the LSD Stokes V profile as weighting.  These average observations were then again converted to LSD profiles, using the same linemasks to redetermine the corresponding $B_l$ values.  In total, we constructed 13 average spectropolarimetric observations from 30 original measurements (see Table\,\ref{tab:phasebin_log}) and kept the four remaining measurements (observations 11, 16, 26, and 27; see Table\,\ref{tab:narval_log}) to adequately cover the full rotation period.  These $B_l$ curves are given in the right panels of Fig.\,\ref{fig:longmag}.  Phase-binning the spectropolarimetric measurements was only possible because of the precise knowledge of the rotation period, determined independently from the magnetic measurements.  As an additional bonus to the increased S/N and smaller uncertainties on the $B_l$ values, the LPVs due to the stellar pulsation modes will be smeared out, since the spectra correspond to different pulsation phases (while having the same rotation phase).  Statistically, one can thus assume that the effects of the LPVs on the phase-binned magnetic measurements are smaller, but it is also possible that we add a systematic distortion or artificially widen the spectral lines.  Yet, the Stokes\,I profiles of the phase-binned data remained rather Gaussian, indicating that most LPVs were indeed cancelled out (see Fig.\,\ref{fig:stokesVmodelling}).  The LSD Stokes\,V profiles of these phase-binned data showed a much clearer Zeeman signature with nine DD and four MD for the full linemask.  Some NDs remain for the other LSD linemasks, because of the lower S/N of these LSD profiles.

We repeated the fitting process of a sine and second order sine function to the determined $B_l$ of these 13 phase-binned and 4 original observations.  Similar results as with the original data were obtained.  The variability agreed with a dipolar field (i.e., a sine function) with a negligible quadrupolar contribution.  The determined fits also agreed well with those of the individual data (see Fig.\,\ref{fig:longmag}).  We do note that some scatter to the modeled $B_l$ variations remained, possibly indicating that not all LPVs are completely cancelled out.

\citet{1987AJ.....94..731S} indicated that the extrema from the dipolar fit to $B_l$ are related through the inclination angle, $i$, and the obliquity angle, $\beta$, through the simple relation:
\begin{equation}
\frac{B_{l,\mathrm{min}}}{B_{l,\mathrm{max}}} = \frac{\cos(i-\beta)}{\cos(i+\beta)}\, \mathrm{.}
\label{eq:shore}
\end{equation}
\noindent Hence, in case one of the angles is constrained, the other is automatically known.  From the dipolar fit to the $B_l$ values of the full linemask, we obtained $B_{l,\mathrm{max}}=272\pm28$\,G and $B_{l,\mathrm{min}}=-184\pm28$\,G.  P12 constrained the inclination angle on HD\,43317 using their derived $P_{\rm rot}$, $v\sin i$, and the radius of an approximate CL{\'E}S model \citep[Code Li{\'e}geois d'{\'E}volution Stellaire;][]{2008Ap+SS.316...83S} to $i \in [20, 50]^{\circ}$.  Our minor update to $P_{\rm rot}$ does not alter this assumption.  Using these limits for the inclination angle and the modulation of the longitudinal magnetic field for the full linemask and no phase-binning, we obtain $\beta \in [69, 88]^{\circ}$.  The estimates for the other models are given in Table\,\ref{tab:longmag}, and these are all compatible with the estimates of B13.

From the fit to the varying longitudinal magnetic field, one can also approximate the strength of the dipolar magnetic field $B_{\rm dip}$.  \citet{1950ApJ...112..222S} derived the analytical equation
\begin{equation}
B_{\rm dip} = \frac{4}{\cos(\beta - i)} \frac{15-5u}{15+u}B_{l,\mathrm{max}} \, \mathrm{,}
\label{eq:strengthdip}
\end{equation}
\noindent where $u$ is the limb-darkening coefficient, which we fix to 0.3, appropriate for a B3.5V star \citep{2000A+A...363.1081C}.  We derived the limits on $B_{\rm dip}$, assuming the inclination angle provided by P12 using the pulsation mode identification, that is, $i \in [20, 50]^{\circ}$, the obtained  $B_{l,\mathrm{max}}$, and the computed $\beta$, with their respective uncertainties, for each of the six different fits.  The derived values are provided in Table\,\ref{tab:longmag}.  Typically, they range from 1 to 2.4\,kG, whenever He lines were included in the linemask.  When no He lines were considered, the field could be slightly weaker (0.6 to 2.2\,kG).  Our results for $B_{\rm dip}$ agree well with those of B13, who deduced 0.65 -- 1.75\,kG and had a slightly smaller value for the amplitude of the sinusoidal fit.

\begin{table*}[t]
\caption{Observing log of the rotationally phase-binned Narval observations, for which the original observation identifications, $\rm ID_{exp}$, from Table\,\ref{tab:narval_log} are indicated.  The rotation phase, $\phi_{\rm rot}$ is determined with $P_{\rm rot} = 0.897673$d and $T_0 = 2456185.8380$\,d.  The provided S/N is that of the LSD Stokes\,I profile calculated with various linemasks.  In addition, the magnetic detection status for each rotation phase-binned observation is indicated (DD = Definite Detection, MD = Marginal Detection, and ND = Non Detection).}
\centering
\tabcolsep=6pt
\begin{tabular}{llccccccc}
\hline
\hline
ID	& $\rm ID_{exp}$	& $\phi_{\rm rot}$	& \multicolumn{2}{c}{all}	&	\multicolumn{2}{c}{He only}	& \multicolumn{2}{c}{He excluded}\\
	&				&				  	& S/N	&Detect.				& S/N	&Detect. 				& S/N	&Detect. \\
\hline											
bin-1	&	4,8		&	0.028952	&	2118	&	MD	&	294	&	MD	&	1735	&	ND	\\
bin-2	&	9,23		&	0.073682	&	2184	&	DD	&	293	&	MD	&	1711	&	MD	\\
bin-3	&	10,24	&	0.118743	&	2151	&	DD	&	291	&	DD	&	1702	&	MD	\\
bin-4	&	3,25		&	0.174297	&	2123	&	DD	&	288	&	DD	&	1689	&	MD	\\
bin-5	&	12,29	&	0.454140	&	2092	&	DD	&	289	&	ND	&	1710	&	ND	\\
bin-6	&	13,17,30&	0.500143	&	2080	&	MD	&	281	&	ND	&	1660	&	MD	\\
bin-7	&	18,31,32&	0.548531	&	2088	&	MD	&	282	&	MD	&	1668	&	ND	\\
bin-8	&	14,19,33&	0.603092	&	2142	&	MD	&	287	&	MD	&	1689	&	ND	\\
bin-9	&	15,34	&	0.652713	&	2095	&	DD	&	285	&	ND	&	1676	&	ND	\\
bin-10	&	1,2		&	0.805048	&	2116	&	DD	&	290	&	DD	&	1687	&	MD	\\
bin-11	&	5,20		&	0.825148	&	2107	&	DD	&	290	&	DD	&	1686	&	MD	\\
bin-12	&	6,21		&	0.870397	&	2059	&	DD	&	293	&	MD	&	1663	&	MD	\\
bin-13	&	7,22		&	0.915811	&	2028	&	DD	&	296	&	MD	&	1630	&	MD	\\
\hline
\end{tabular}
\label{tab:phasebin_log}
\end{table*}

\section{Stokes\,V modeling}
\label{sec:vmodelling}
In an attempt to constrain the geometry of the large-scale magnetic field of HD\,43317 more accurately, especially the inclination and obliquity angles, we resorted to modeling the Zeeman signatures with a grid-based approach \citep[for full details, see][]{2008MNRAS.385..391A}.  For each observation, a (de-)centered dipolar magnetic field is described and characterized by $i$, $\beta$, the strength of the dipolar field $B_{\rm dip}$, a reference date $T_0$ for the rotation phase of the maximal amplitude of $B_{\rm dip}$, and the off-centering factor of the field $\psi$.  The latter varies between 0 and 1, with 0 being a centered dipolar field.  For each gridpoint in this five-dimensional grid, the Stokes\,V profiles were integrated over the $(\theta, \phi)$ stellar surface mesh.  The routine, then, deduced the best solution by simultaneously minimizing the reduced chi-square for every observation to their respective Stokes\,V model within the velocity range of $[-350,350]$\,km/s.

To accurately represent the linewidth of the modeled profiles, we provided the assumed limb-darkening coefficient $u=0.3$, and a description of the LSD Stokes\,I profile, parametrized by the radial velocity shift, the depth of the profile, the rotational broadening $v\sin i$, and the non-rotational broadening $v_{\rm NR}$, to the modeling tool.  The description was obtained by performing a LS minimization fit of the convolution product of two Gaussian functions to the observed LSD Stokes\,I profiles, for which one Gaussian had a fixed full width at half maximum (FWHM) equal to the literature value for $v\sin i$ derived by P12, and the other Gaussian has $v_{\rm NR}$ as a fitting parameter for its FWHM.  This indicated $v_{\rm NR} = 110$\,km/s, which was not surprising given the wide LSD Stokes\,I profiles due to various broadening effects, coming from the different included spectroscopic lines.  This non-rotational broadening comprises of, at least, the macroturbulent broadening, the pressure broadening of the He lines, and the combined amplitudes of the stellar pulsation modes, seen at various optical depths (through the different spectroscopic lines).

Because the S/N in Stokes\,V for the phase-binned data is substantially higher, we performed the Stokes\,V modeling to this dataset.  Moreover, we only modeled the LSD profiles constructed with the complete linemask, since i) it has the highest S/N of the LSD profiles constructed with the different linemasks, ii) the I-profiles of the He-only linemask start to differ from a strict Gaussian profile (see Fig.\,\ref{fig:stokesVoHe}), iii) no DD were obtained with the He-excluded linemask, and iv) the effects of the LPVs should be minimal.  

\subsection{Full parameter search}
\label{sec:freevmodel}
Because we did not want to enforce any values for the parameters or restrict the parameter space, we left $i$ and $\beta$ to vary between 0 and 90\,$^{\circ}$, $B_{\rm dip}$ between 100 and 2500\,G, and $T_0$ in the range of 2456185\,d and 2456185+$P_{\rm rot}$\,d.  We compare the final, returned model with a centered dipole ($\psi=0$) to the LSD profiles in Fig.\,\ref{fig:stokesVmodelling}.  These models indicate $i=55.6 \pm 6.4\,^{\circ}$, $\beta=77.0\pm 4.5\,^{\circ}$ with $B_{\rm dip} = 982\pm45$\,G and $T_0 = 2456185.8380 \pm 0.0059$\,d.  The provided uncertainties here are $3\sigma$ confidence intervals, compared to the $68\,\%$ ($1\sigma$) confidence intervals throughout the work.  When setting $\psi$ as a free parameter during the Stokes\,V modeling, the models with $\psi=0$ described the observations most accurately, hence a centered dipolar magnetic field was always favored.   This result is compatible with those of the longitudinal magnetic field study, indicating no significant quadrupolar contribution to the dipolar magnetic field.

Comparing the modeled Stokes\,V profiles with those observed, we noted that the agreement is good but not perfect (a reduced $\chi^2=1.13$).  While the rotational variability of the Zeeman signature is adequately described, differences compared to the individual observations remain visible, yet fall within the uncertainty on the observed LSD Stokes\,V profiles.  These differences seem to be more substantial when the magnetic field was not observed close to pole-on.  In some cases, differences most likely arose because of the considerable noise level of these observations, reflected by the variable Null profile.  Other observations showed that a minor distortion by the LPVs still occurred when comparing the intensity profiles with the Gaussian approximation.  The differences between the Stokes\,V model and the polarization signature increases whenever the I-profile differs from the anticipated Gaussian approximation (for example at $\phi_{\rm rot} = 0.6527$ or $\phi_{\rm rot} = 0.8704$).  We, therefore, assume that the combination of both aforementioned issues causes the differences between models and observations.

\subsection{Restricted parameter search}
\label{sec:restrictedvmodel}
The inclination angle determined from the Stokes\,V modeling differs from the seismic estimate of P12.  The grid-based approach indicates $55.6\pm6.4^{\circ}$, which is only marginally compatible with the limits derived by P12, that is, $i \in [20, 50]^{\circ}$.  Consequently, this larger inclination angle would imply a lower equatorial velocity, when keeping $v\sin i$ constant.  Hence, to obtain the same rotation period, HD\,43317 needs to have a substantially smaller stellar radius, of about $2.5\,\mathrm{R_{\odot}}$.  Such a radius is much smaller than what is typically determined and expected for a B3IV star (i.e., about 4\,$\mathrm{R_{\odot}}$).  

Therefore, we repeated the Stokes\,V modeling with the inclination angle fixed at $30\,^{\circ}$, close to the estimate of P12.  The resulting parameters for the determined model (with $i$, $\psi$, and $T_0$ fixed) were $\beta = 85 \pm 2\,^{\circ}$ and $B_{\rm dip} = 1100 \pm 43$\,G.  $\beta$ is now artificially better constrained because the correlation between $i$ and $\beta$ was not accounted for by keeping $i$ fixed.  A larger $B_{\rm dip}$ was needed to correctly model the amplitudes of the Zeeman signature.  This additional model is indicated in Fig.\,\ref{fig:stokesVmodelling} for a visual comparison with the model of Sect.\,\ref{sec:freevmodel}.

The resulting reduced chi-square of the new model, $\chi^2 = 1.15$, is only slightly worse than the model obtained in Sect.\,\ref{sec:freevmodel}.  Hence, we were unable to reliably constrain $i$ from Stokes\,V modeling.  The LPVs in combination with the mediocre S/N are most likely providing an overly large challenge for the standard techniques.  In case HD\,43317 is seen from a low inclination angle, as anticipated, the star is rotating at a significant fraction of its critical rotation velocity.  This could provide an additional observational obstacle for the study of the spectropolarimetric data.  Fortunately, we did not note any evidence in the I-profiles of a strong oblate stellar deformation.
\section{Discussion}
\label{sec:discussion}
\subsection{Magnetic dipole configuration}
\label{sec:inclinationdiscussion}
Using both the fit to the varying longitudinal magnetic field, together with the corresponding analytical equations, and the Stokes\,V modeling, we have tried to characterize the large-scale magnetic field at the surface of HD\,43317 as a dipolar magnetic field and have attempted to determine the inclination angle of the stellar axis with respect to the observer.  The applied methods had varying degrees of success, but always favored a dipolar geometry for the large-scale magnetic field.  A minor quadrupolar contribution to the dipolar field is still plausible, but not needed with the current spectropolarimetric dataset and accepted rotation period.  Hence, we continue to treat the large-scale magnetic field at the stellar surface of HD\,43317 as a pure dipolar magnetic field.

The Stokes\,V modeling, often a strong technique to constrain $i$, was not able to reliably determine the inclination angle of HD\,43317, since the best fit led to a non-physical stellar radius for a B3.5V star.  Moreover, enforcing the inclination angle from P12 during the modeling resulted in only a slightly worse description of the Zeeman signatures.  Therefore, to reliably constrain the inclination angle, we will need to follow a seismic approach by modeling the observed stellar pulsation-mode frequencies, while assuming the pulsation axis coincides with the rotation axis.  This is beyond the scope of the current paper.

Although $i$ can span a wide range of values, we have a relatively good handle on the obliquity angle between the magnetic axis and the rotation axis.  Both the Stokes\,V modeling and Eq.\,(\ref{eq:shore}) indicated that $\beta$ should be large ($>67\,^{\circ}$), yet be smaller than $90\,^{\circ}$.  A similar conclusion was reached for the location of the co-rotating He patches (i.e., $\beta_{\rm spot} \in [60, 80]^{\circ}$).  Therefore, it seems that these two surface abundance inhomogeneities are located very close to the magnetic poles, as is often found for magnetic stars.

Both $i$ and $\beta$ will play a fundamental role in correctly assessing and investigating the possibility of magnetic splitting of the stellar pulsation-mode frequencies seen in the CoRoT photometry, since they directly regulate the multiplet structure \citep[e.g.,][]{1982MNRAS.201..619B, 1983MNRAS.205.1171R, 1993PASJ...45..617S, 1996ApJ...458..338D, 2000A+A...356..218B, 2005A+A...444L..29H}.

Lastly, from the longitudinal field measurements, we obtained $B_{\rm dip}$ ranging from 1 to 2.4\,kG, while the Stokes\,V modeling indicated $\sim\,1$\,kG.  It is not the first time that Stokes\,V modeling provides a lower amplitude for the field.  In the particular case for HD\,43317, we note at least two reasons.  First, the value derived from the Stokes\,V modeling relies heavily on the considered value for the non-rotational broadening $v_{\rm NR}$, where a larger $v_{\rm NR}$ corresponds to a larger $B_{\rm dip}$. Accurately deriving $v_{\rm NR}$ is not trivial for early-type stars that host stellar pulsation modes or show rotational modulation \citep[see e.g.,][]{2014A+A...569A.118A}.  As such, the uncertainty on $B_{\rm dip}$ is larger than what the Stokes\,V modeling has indicated.  Second, the upper limits for $B_{\rm dip}$ derived from the longitudinal magnetic field (i.e., $> 2$\,kG) occurs only under the assumption of the low inclination angle limit, that is, $i = 20\,^{\circ}$, and thus comes from the uncertainty on $i$ derived by P12.  Using the averaged longitudinal field measurements for the complete LSD linemask (top right panel of Fig.\,\ref{fig:longmag}) and assuming $i=30\,^{\circ}$ for Eq.\,(\ref{eq:strengthdip}), we obtain a $B_{\rm dip}=1.4$\,kG, which is closer to the Stokes\,V modeling results.  Because of the uncertain $v_{\rm NR}$, the large interval for $i$, and the fast rotation, we, therefore, conclude that the magnetic dipolar field at the stellar surface of HD\,43317 likely has a strength between approximately 1 and 1.5\,kG.

\subsection{Circumstellar environment}
\begin{table}[t]
\caption{Derived values needed for the study of the magnetosphere of HD\,43317, assuming two different inclination angles.  Both computations indicate HD\,43317 hosts a centrifugal magnetosphere. }
\centering
\tabcolsep=4pt
\begin{tabular}{lcc}
\hline
								&	$i=20\,^{\circ}$		 & $i = 50\,^{\circ}$	\\
\hline
$R_*$ ($R_{\odot}$)				& 5.966					& 2.664					\\
$\log \dot{M}$ ($M_{\odot}/yr$)	& -8.945					& -10.493				\\
$\eta_*$ 						& 130.5					& 7367.8					\\
$R_{\rm A}$ ($R_{\odot}$)		& 21.964					& 25.48				\\
$R_{\rm K}$ ($R_{\odot}$)		& 6.87					& 6.87					\\
\hline
\end{tabular}
\label{tab:magnetosphere}
\end{table}

The presence of a magnetosphere around magnetic stars is often noted by the rotationally modulated H$\alpha$ emission.  No such variability was found from the new Narval data, in agreement with the conclusions of both P12 and B13, using the HARPS spectroscopy and the 2011--2 Narval spectropolarimetry, respectively.  This, however, does not necessarily imply that a magnetosphere is absent for HD\,43317, since the strength of the variability caused by the magnetosphere depends on the viewing angle upon it.  Under the oblique rotator model \citep{1950MNRAS.110..395S}, this viewing angle depends on $i$, $\beta$, and the rotation phase, resulting in limited variability when either the inclination angle or the obliquity angle is small ($< 20\,^{\circ}$) while the other one is large ($>70\,^{\circ}$).  A similar conclusion was reached by \citet{2008MNRAS.389..559T}, when studying the effect of the viewing angle upon the photometric variability when using a torus model for the magnetosphere.  This unfavorable geometrical configuration is compatible with the current confidence intervals on $i$ and $\beta$ for HD\,43317.

To determine whether HD\,43317 hosts a dynamical or centrifugal magnetosphere, we determine the magnetic confinement parameter $\eta_*$ \citep[using Eq.\,(7) of][]{2002ApJ...576..413U}, considering the adopted 1\,kG dipolar magnetic field and the escape velocity $v_{\rm \infty} = 150$\,km/s from B13.  We computed the theoretical mass-loss rate $\dot{M}$ using Eq.\,(25) of \citet{2001A+A...369..574V}, because no accurate observed $\dot{M}$ is available for HD\,43317.  To be compatible with the large confidence interval on the inclination angle, we performed the calculations for two different inclination angles, namely $i=20\,^{\circ}$ and $i = 50\,^{\circ}$, leading to different stellar radii, while assuming the literature $v\sin i$ and the adopted rotation period.  Using the computed $\eta_*$, we determined the Alfven radius $R_{\rm A}$ for both cases.  The Keplerian corotation radius $R_{\rm K}$ was derived assuming HD\,43317 has a mass of $M = 5.4\,M_{\odot}$, following the CL{\'E}S model of P12, and the adopted rotation period.  These values for the different parameters are given in Table\,\ref{tab:magnetosphere}.  In both cases, we obtained $R_{\rm A} > R_{\rm K}$, indicating HD\,43317 hosts a centrifugal magnetosphere, compatible with the results of B13.  This result remains valid when HD\,43317 has a stronger dipolar magnetic field than the assumed 1\,kG.

\subsection{Implications for the stellar interior}
Since HD\,43317 is a magnetic early-type pulsator with many detected stellar pulsation-mode frequencies, it will be a prime target to observationally inspect the implications of the large-scale magnetic field for the stellar interior.  In a future paper, the observed pulsation modes will be used to constrain the stellar structure through forward modeling \citep[as done in, e.g.,][]{2010Natur.464..259D, 2015ApJ...803L..25P, 2016A+A...593A.120V, 2016ApJ...823..130M}.  At this stage, however, we can already employ several theoretical criteria to study whether or not the magnetic field at the stellar surface of HD\,43317 is sufficiently strong to imply solid-body rotation.

The first criterion is Eq.\,(3.6) of \citet{2011IAUS..272...14Z}, based on the magnetic field description by \cite{2005A+A...440..653M}:
\begin{equation}
B^2_{\rm crit} = 4 \pi \rho \frac{R^2_*\Omega}{\tau_{\rm AM}}\,\mathrm{.}
\label{eq:zahncriterion}
\end{equation}
\noindent Here, $R_*$, $\rho$, and $\Omega$ are the stellar radius, the mean density, and the surface velocity, respectively.  $\tau_{\rm AM}$ is the time spent by the star on the main sequence.  Until better estimates have been derived, we used the radius and mass of the CL{\'E}S model of P12 to estimate $\rho$.  To calculate the surface velocity, we adopted $i=30\,^{\circ}$ and $v\sin i = 115$\,km/s, again following P12.  For the age of HD\,43317, we employed the result determined by \citet[][]{2011MNRAS.410..190T}, who derived $\tau_{\rm AM} = 26.0 \pm 5.6$\,Myr by fitting HD\,43317 to various evolutionary models.  This lead to a $B_{\rm crit}\approx 114$\,G in the radiative zone to suppress differential rotation.  Employing the ratio between the internal and surface magnetic fields from \citet{2008MNRAS.386.1947B}, that is, a ratio of 30, we obtained a $B_{\rm crit, surf} \approx 4$\,G.  The derived magnetic field of HD\,43317 is substantially stronger than this limit, implying that under the Zahn-criterion, the magnetic field will influence the internal stellar structure by enforcing uniform rotation and inhibiting mixing.  In case HD\,43317 is older than the assumed 26\,Myr, the derived $B_{\rm crit, surf}$ will decrease even further.  Moreover, HD\,43317 needs to be younger than about 1\,Myr, conflicting with the results of \citet[][]{2011MNRAS.410..190T}, to significantly challenge the derived $B_{\rm crit, surf}$ and its implications.

An alternative criterion was derived by \citet{1999A+A...349..189S}, which calculates the minimal initial field strength for uniform rotation when assuming a non-axisymmetric magnetic field:
\begin{equation}
B_{\rm crit, init} = r (4\pi \rho)^{1/2}\left(\frac{\eta \Omega q^2}{3 r^2 \pi^2}\right)^{1/3}\,\mathrm{,}
\label{eq:spruitcriterion}
\end{equation}
\noindent while only considering the radiative zone.  We used similar approximations to those of \citet[][for the magnetic pulsator V2052\,Oph]{2012MNRAS.427..483B}, and assumed $i=30^{\circ}$.  This leads to a $B_{\rm crit, init} \approx 1400$\,G, corresponding to a $B_{\rm crit, init, surf}\approx 140$\,G using the ratio (10) between the surface field and that in the radiative zone \citep{2008MNRAS.386.1947B}.  Again, the measured dipolar field at the surface of HD\,43317 passes this threshold field, implying uniform rotation.

This anticipated uniformly rotating radiative layer will cause a decrease in the size of the convective core overshooting layer, leading to less available fuel for the core-hydrogen-burning phase compared to a non-magnetic star with the same stellar parameters as HD\,43317.  From forward seismic modeling, it should be possible to observationally parametrize this overshooting layer.

Since the inclination angle is currently ill-defined, we investigate what happens to the critical values if we assume $i=55.6\,^{\circ}$, following the result of the Stokes\,V modeling, while keeping $R_*$ constant (conflicting with the observed $v\sin i$) or while keeping $v\sin i$ constant (leading to the non-physical $R_* = 2.5\,R_{\odot}$.  In the first case, both critical values would decrease, due to the slower surface velocity, to $B_{\rm crit, surf}=3$\,G and $B_{\rm crit, init, surf} \approx 100$\,G.  The inverse behavior is noted for the second case, as the critical values increase to $B_{\rm crit, surf}=5.5$\,G and $B_{\rm crit, init, surf} \approx 310$\,G.  Nevertheless, the determined polar strength between approximately 1 and 1.5\,kG for the fossil field of HD\,43317 remains sufficiently large, compared to the derived critical values to warrant a uniformly rotating radiative envelope.

\section{Conclusions}
\label{sec:conclusions}
Using archival and recent Narval spectropolarimetry, we studied the large-scale magnetic field of HD\,43317 and the LPVs, due to stellar pulsation modes and co-rotating surface abundance inhomogeneities.

Thanks to the carefully tailored exposure times of the individual polarization sub-exposures, variations of the line profiles during the complete sequences were minimal.  However, the LPVs cause periodic distortions of the spectral lines over longer timescales.  By combining the individual Narval sequences with the HARPS spectroscopy, assuming two surface abundance inhomogoneities, and fitting the variability of the first moment of the \ion{He}{I}\,6678\,$\AA$ line, we updated the rotation period of HD\,43317 to $P_{\rm rot} = 0.897673 \pm 0.000004$\,d.  Our value for the rotation period is compatible with the value determined by P12, using the CoRoT photometry.

Employing the LSD technique, we constructed mean intensity and polarization line profiles for each Narval sequence using three linemasks.  The rotational modulation of the derived values for the longitudinal magnetic field agreed with a dipolar magnetic field and did not indicate any substantial quadrupolar contribution.  Thanks to the precisely derived rotation period, we were able to bin and average spectropolarimetric observations with a similar rotation phase.  The longitudinal field measurements of these phase-binned data also agreed with a dipolar magnetic field.  Small differences are seen for the derived model and may be caused by the LPVs due to the stellar pulsation modes.  The results obtained from the various characterizations agree with those derived by B13.

We also attempted to characterize the geometry of the dipolar field by comparing each LSD profile of the phase-binned Narval spectropolarimetry to a grid of synthetic polarization profiles produced by a (de)centered dipolar magnetic field at the stellar surface.  The resulting description of the large-scale magnetic field represented the variability of the Zeeman signature with the rotation phase relatively well, but failed to accurately characterize each observed Stokes\,V profile.  The differences are comparable with the noise level of the LSD Stokes\,V profiles.  We assume that these differences were due to a combination of the mediocre S/N level, the rather small variability of the Stokes\,V profiles, and the LPVs distorting the I-profiles.  Moreover, the derived inclination angle ($i=55.6 \pm6.4\,^{\circ}$) from the model led to an overly small stellar radius for HD\,43317.  An alternative model with the inclination angle fixed to the value of P12 ($i=30\,^{\circ}$) resulted in only a slightly worse description of the observed polarization profiles.  Hence, constraining the inclination angle from the spectropolarimetric data only seems to be difficult.  We intend to resolve this problem with a detailed and coherent magneto-asteroseismic analysis in a following paper.  The obliquity angle of the magnetic field to the rotation axis seems to be substantial, but less than $90\,^{\circ}$, irrespective of the technique employed.  Similar constraints were obtained for the co-rotating He patches at the stellar surface of HD\,43317, placing them close to the magnetic poles.

We deduced that the dipolar magnetic field of HD\,43317 is between approximately 1 and 1.5\,kG from the analytical equations for the longitudinal magnetic field measurements and from the Stokes\,V modeling.  Small discrepancies for the field strength between the two methods occur due to the uncertain inclination angle and the uncertain line broadening.  Comparing the derived strength for the magnetic field with theoretical criteria indicates that the surface magnetic field of HD\,43317 is sufficiently strong to influence the structure of the stellar interior by imposing uniform rotation in the radiative layer of the star.  We intend to confirm these implications by performing forward seismic modeling.

\begin{acknowledgements}
This work has made use of the VALD database, operated at Uppsala University, the Institute of Astronomy RAS in Moscow, and the University of Vienna, of the SIMBAD database operated at CDS, Strasbourg (France), and of NASA's Astrophysics Data System (ADS). 

Part of the research leading to these results has received funding from the European Research Council (ERC) under the European Union’s Horizon 2020 research and innovation programme (grant agreement N$^\circ$670519: MAMSIE).

\end{acknowledgements}
\bibliographystyle{aa}
\bibliography{PhD_ADS}

\begin{appendix}
\section{LSD profiles}
\label{sec:appendix_LSD}
In this section, we provide all LSD profiles of each studied Narval spectropolarimetric sequence.  Each figure represents one of the considered linemasks.
\begin{figure*}[th]
		\centering
			\includegraphics[width=\textwidth, height = 0.90\textheight]{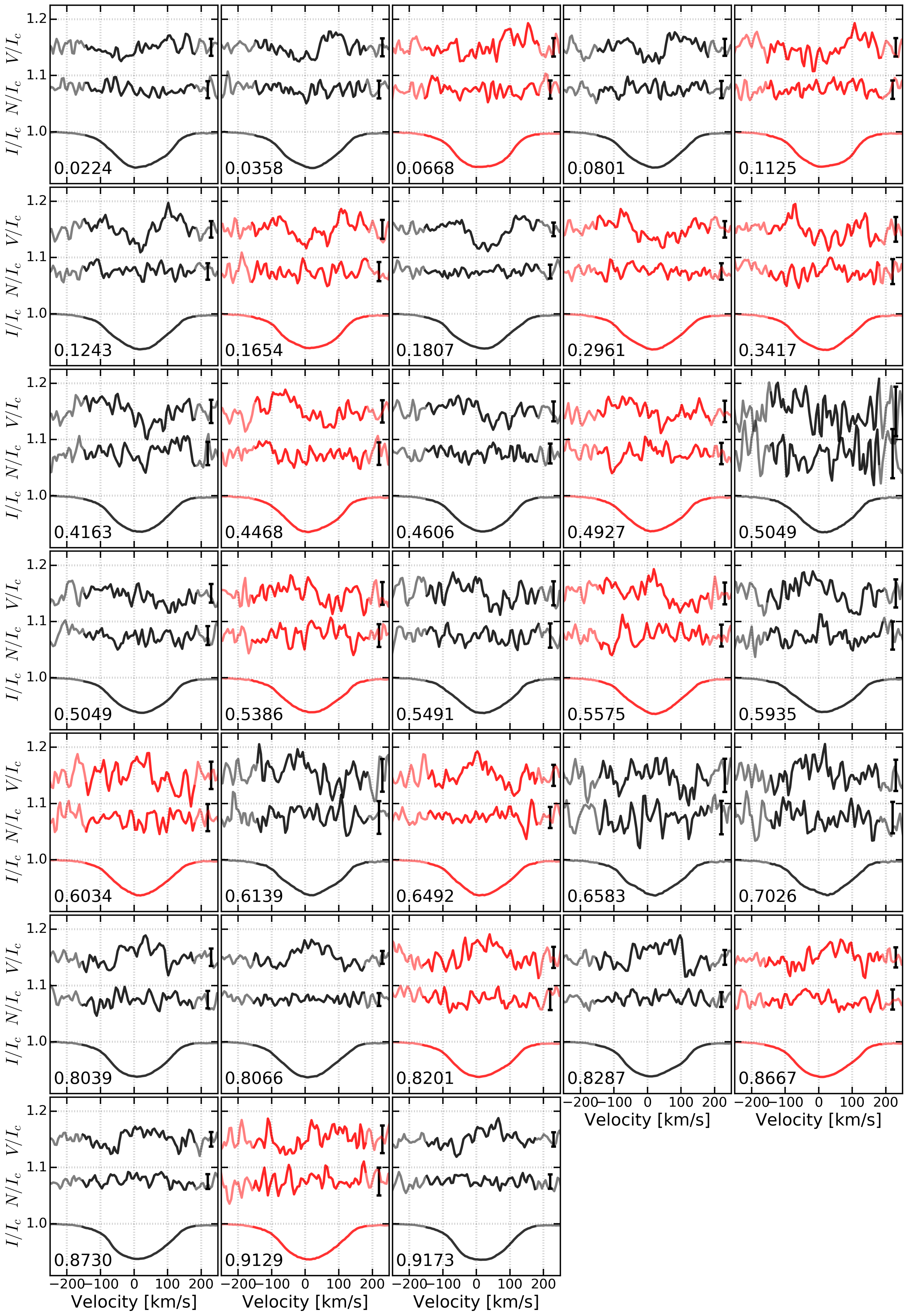}%
			\caption{LSD profiles determined with the full linemask for HD\,43317 (Sect.\,\ref{sec:magnetic}).  Each panel is organized according to the rotation phase and contains the observed intensity profile ($I/I_c$; \textit{bottom}), the Null profile ($N/I_c$; \textit{middle}), and the Stokes\,V ($V/I_c$; \textit{top}) profiles.  Both the Null and Stokes\,V profiles are multiplied by a factor 100 and smoothed with a Gaussian filter with width 6\,km/s for increased visibility.  Their respective error bars are indicated on the right.  Narval data discussed in B13 is marked in black, and the more recent Narval spectropolarimetry presented in this work is indicated in red.}
			\label{fig:stokesVall}
\end{figure*}
\begin{figure*}[th]
		\centering
			\includegraphics[width=\textwidth, height = 0.90\textheight]{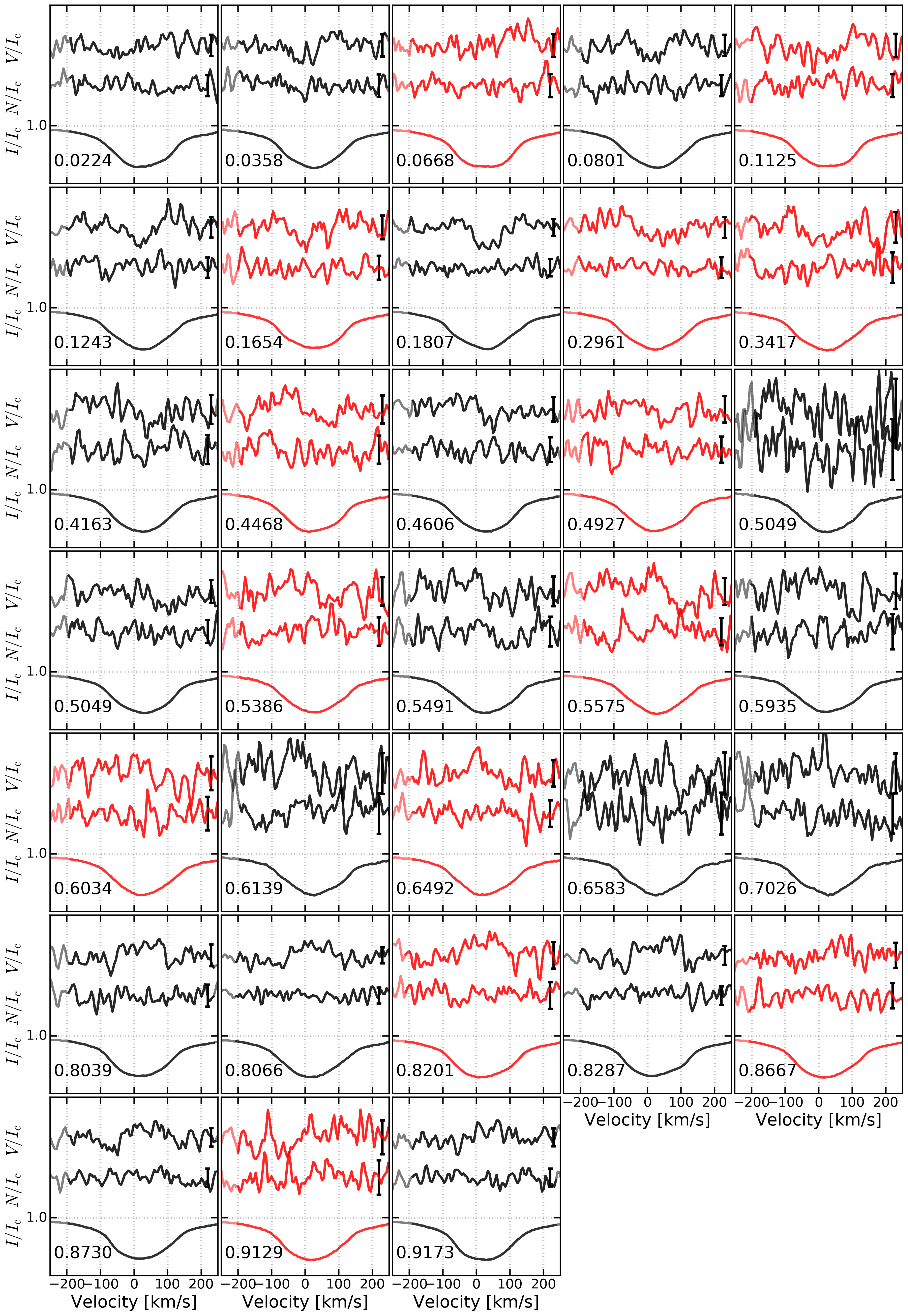}%
			\caption{As in Fig.\,\ref{fig:stokesVall}, but for the LSD profiles determined with only the 17 He lines for HD\,43317.}
			\label{fig:stokesVoHe}
\end{figure*}
\begin{figure*}[th]
		\centering
			\includegraphics[width=\textwidth, height = 0.90\textheight]{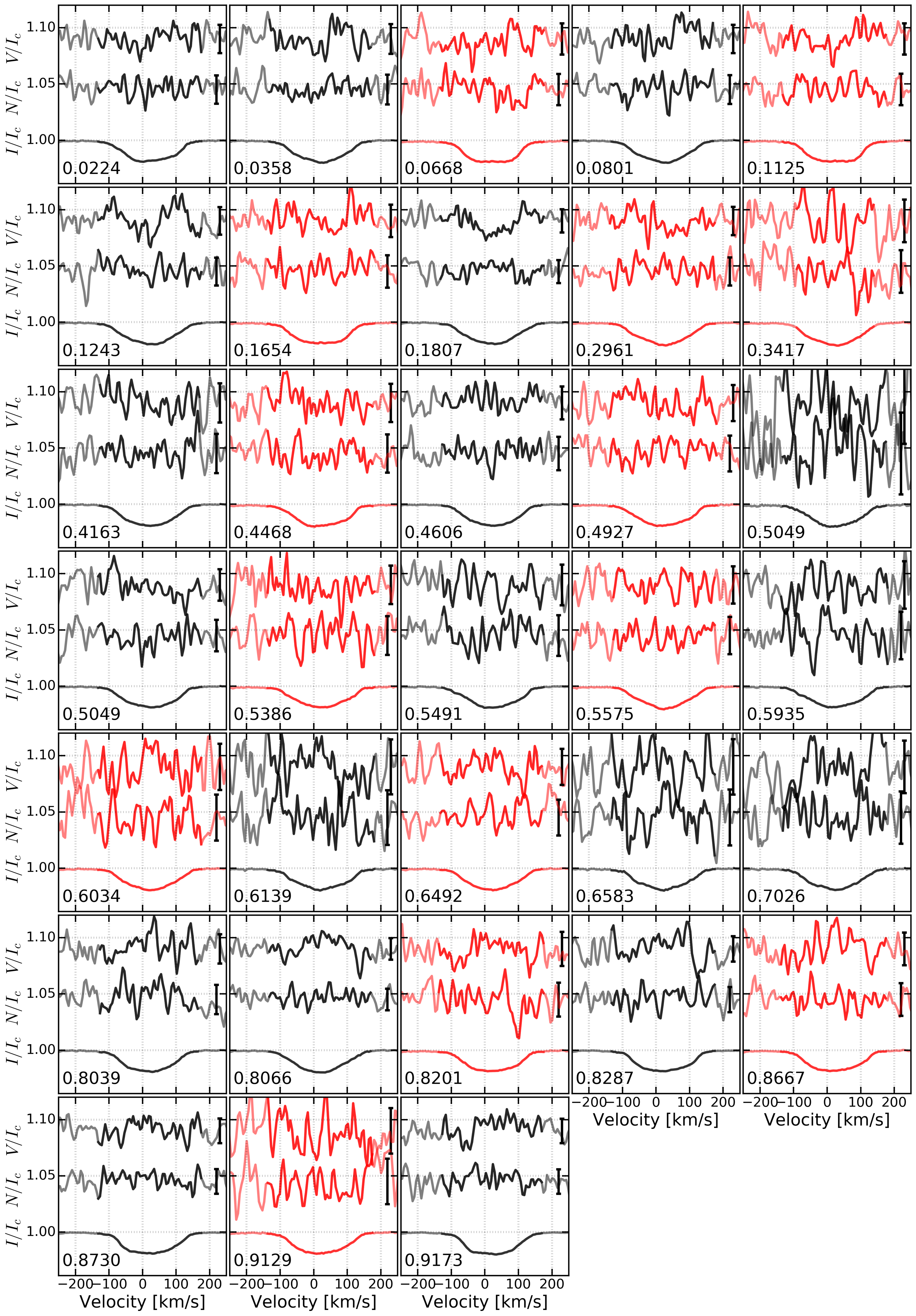}%
			\caption{As in Fig.\,\ref{fig:stokesVall}, but for the LSD profiles determined with all significant metallic lines except the (blends with) He lines for HD\,43317.}
			\label{fig:stokesVnHe}
\end{figure*}
\end{appendix}
\end{document}